\documentclass[journal]{IEEEtran}
\usepackage{cite}
\usepackage{amsmath,amssymb,amsfonts}
\usepackage{algorithmic}
\usepackage{graphicx}
\usepackage{mathtools}
\usepackage{color,soul}
\usepackage{epstopdf}
\usepackage{graphics}
\usepackage{amsmath}
\usepackage{algorithm}
\usepackage{amsfonts,amssymb,amsthm}
\usepackage{upgreek}
\usepackage[tight]{subfigure}
\usepackage{multirow}
\usepackage{tabularx}
\usepackage{dblfloatfix}    
\usepackage[normalem]{ulem}
\newcommand{\noindento}{\vspace{0.1cm}\noindent} 

\usepackage{acronym}

\acrodef{RIS}{Reconfigurable Intelligent Surface}
\acrodef{LIS}{Large Intelligent Surface}
\acrodef{SDM}{Software-Defined Metamaterial}

\begin{document}
\title{Scalability Analysis of Programmable Metasurfaces for Beam Steering}
\author{
Hamidreza Taghvaee,
Sergi Abadal,
Alexandros Pitilakis,
Odysseas Tsilipakos,
Anna Tasolamprou,
Christos K Liaskos,
Maria Kafesaki,
Nikolaos V. Kantartzis,
Albert Cabellos-Aparicio,
and Eduard Alarc\'{o}n%
\thanks{H. Taghvaee, S. Abadal, A. Cabellos-Aparicio, and Eduard Alarcón are with the NaNoNetworking Center in Catalonia (N3Cat), Universitat Polit\`{e}cnica de Catalunya, 08034 Barcelona, Spain (corresponding e-mail: taghvaee@ac.upc.edu)}
\thanks{A. Pitilakis and N. V. Kantartzis are with the Department of Electrical and Computer Engineering, Aristotle University of Thessaloniki, Thessaloniki, Greece.}
\thanks{O. Tsilipakos, A. Tasolamprou, C. K. Liaskos and M. Kafesaki are with the Foundation for Research and Technology Hellas, 71110, Heraklion, Crete, Greece.}
}


\maketitle

\begin{abstract}
Programmable metasurfaces have garnered significant attention as they confer unprecedented control over the electromagnetic response of any surface. Such feature has given rise to novel design paradigms such as Software-Defined Metamaterials (SDM) and Reconfigurable Intelligent Surfaces (RIS) with multiple groundbreaking applications. However, the development of programmable metasurfaces tailored to the particularities of a potentially broad application pool becomes a daunting task because the design space becomes remarkably large. 
This paper aims to ease the design process by proposing a methodology that, through a semi-analytical model of the metasurface response, allows to derive performance scaling trends as functions of a representative set of design variables. Although the methodology is amenable to any electromagnetic functionality, this paper explores its use for the case of beam steering at 26 GHz for 5G applications. Conventional beam steering metrics are evaluated as functions of the unit cell size, number of unit cell states, and metasurface size for different incidence and reflection angles. It is shown that metasurfaces 5$\lambda\times$5$\lambda$ or larger with unit cells of $\lambda/3$ and four unit cell states ensure good performance overall. Further, it is demonstrated that performance degrades significantly for angles larger than $\theta > 60^o$ and that, to combat this, extra effort is needed in the development of the unit cell. These performance trends, when combined with power and cost models, will pave the way to optimal metasurface dimensioning.

\end{abstract}


\section{Introduction}
\label{sec:intro}

The fifth generation (5G) of mobile communications is sustained by a set of key technologies that allow to satisfy the increasing speed, efficiency, and connectivity demands of wireless networks \cite{akyildiz20165g}. Relevant examples are massive MIMO~\cite{bjornson2016massive}, millimeter-wave spectrum use \cite{rappaport2013millimeter}, or software-defined networking \cite{farhady2015software}. However, a large body of research is already focusing on the major challenges and opportunities to shape the \emph{sixth generation} of wireless networks~\cite{bjornson2019massive, nawaz2019quantum, Liaskos2018a, nie2019intelligent, petrov2019exploiting, mestres2017knowledge}.

In this context, the concept of Software-Defined Metasurfaces (SDMs) has garnered considerable attention as they allow to modify at will the characteristics of the waves that impinge on it~\cite{AbadalACCESS,SDM2015design,Pitilakis2018MMParadigm}. Using SDMs or other variants of the concept such as Reconfigurable Intelligent Surfaces (RIS), wireless environments become programmable and can be incorporated within the design loop of the network [Fig. 1(a)]. This represents a true paradigm shift in wireless networks, where the channel has traditionally been an inevitable limiting factor, and opens the door a plethora of novel co-design techniques with enormous potential as the recent explosion of works can attest~\cite{Liaskos2018,di2019smart,di2019reflection,tan2018enabling,huang2019reconfigurable,tang2019wireless,dai2019wireless,arun2019rfocus}.     


Programmable metasurfaces (MS) are the key enablers of the SDM/RIS paradigm. MSs are compact and planar arrays of subwavelength controllable resonators, i.e., the unit cells. The subwavelength granularity of these \emph{unit cells} confers MSs with exceptional control of electromagnetic (EM) waves as demonstrated in a variety of works \cite{Munk, Glybovski2016, 6230714, Koschny2017,Yang2016a,Tasolamprou201423147, Li2017b, Tcvetkova2018, Tsilipakos:2018aom, Wang2018, Qu2015, Zhang2016a, Liu2016a, Perrakis:2019, Hosseininejad2019a}. The actual response of the MS is derived from the aggregated response of all unit cells, which need to be modified individually. For instance, beam steering requires exerting specific amplitude and phase profiles to the impinging wave \cite{Yu2011a,Pfeiffer:2013,Tsilipakos:2018aom,Liu:2019, Hosseininejad2019a}.


\begin{figure*}[!t]
\centering
\vspace{-0.5cm}
\includegraphics[width=0.9\textwidth]{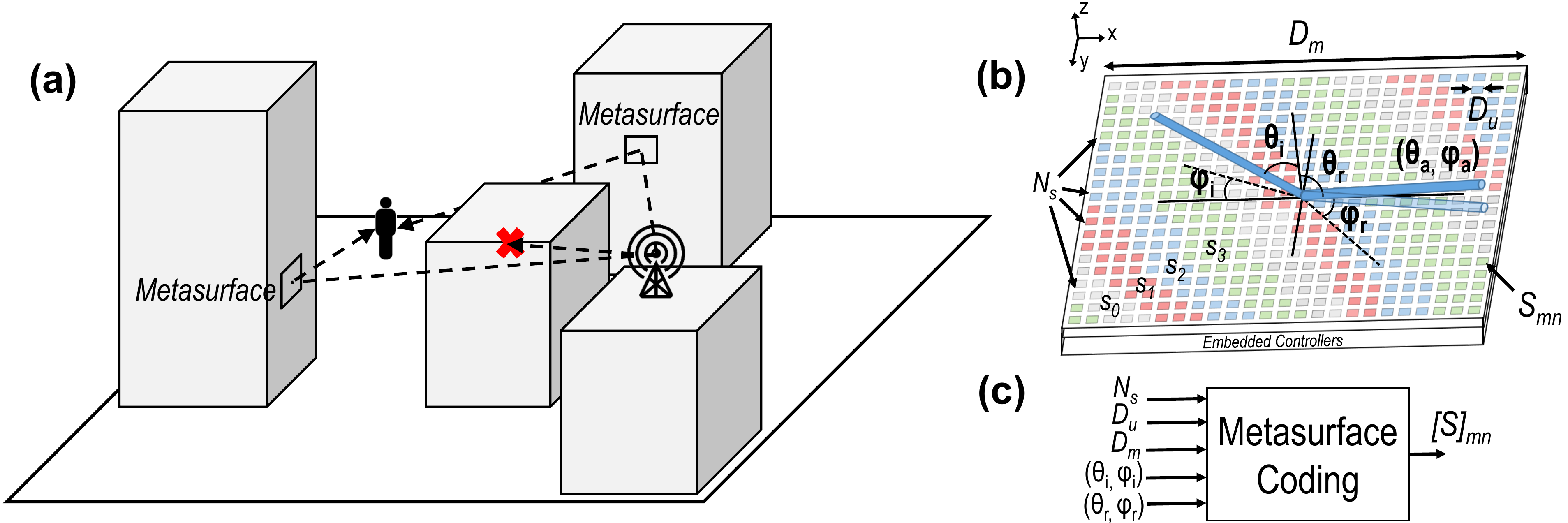} 
\vspace{-0.1cm}
\caption{Schematic representation of (a) a wireless environment augmented with programmable metasurfaces for coherent combination of reflected rays, (b) a metasurface of size $D_m$ for beam steering with unit cells of size $D_u$ and $N_s$ possible states ($s_0$, $s_1$, $s_2$, $s_3$), and (c) the process of metasurface coding.}
\label{Mdl}
\end{figure*}

Programmability in MSs is achieved via the inclusion of tunable elements within the MS structure and the addition of means of control over such tunable elements~\cite{Oliveri2015, Makarov2017, Liu2018ISCAS, Liu:2019, Cui2019, abadal2020programmable}. These aspects have led to the recent proposal of MSs that could be indeed \emph{encoded}, this is, where the polarization-phase-direction of the reflected beam can be controlled by (re)programming each single cell unit choosing among a finite set of states \cite{Cui2014}. At the hardware level, this has been implemented either by using external Field-Programmable Gate Arrays (FPGAs) \cite{Zhang2018b} or by directly embedding the controllers within the MS structure~\cite{AbadalACCESS, Liaskos2018,Mehrotra2019, Tasolamprou8788546,Liaskos2019}. At the software level, the encoding process can be tackled by modeling the EM functionalities via a set of well-defined software primitives~\cite{HsfNetworkTNET.2019}.   



The promises of the SDM/RIS paradigm, however, come at the expense of a non-trivial complexity in the MS. On the one hand, the performance of a SDM depends on the size of the unit cells, the number of unit cell states, or the size of the whole MS. On the other hand, there are costs and energy overheads associated with the fabrication and operation of SDMs that also scale with the aforementioned factors \cite{abadal2020programmable}. Hence, in order build SDMs capable of satisfying a set of application-specific requirements with the minimum cost, it becomes necessary to quantify the main scaling trends and tradeoffs of the underlying MS.



This paper aims to bridge this gap by providing a method to dimension the SDM/RIS through a design-oriented scalability analysis of programmable MSs. In particular, we study the impact of relevant design parameters on the potential performance of programmable MS. Coupled to power consumption, cost, or application-specific models, our methodology will provide SDM/RIS designers and network architects with a clear picture of the practicable design space, illustrating the main tradeoffs and pointing to potentially optimal regions. Although programmable MSs have been the subject of sensitivity analyses \cite{Moccia2017a, taghvaee2020error}, the impact of scaling fundamental design parameters has not been studied yet. Bj\"{o}rnson \emph{et al.} studied the scaling of power in RIS environments, but considers conventional arrays rather than programmable MSs \cite{bjornson2020power}.



The main contributions of this paper are threefold. First, we declare a general design-oriented and model-based methodology to perform a scalability analysis of programmable MSs. Second, and although the methodology is amenable to any functionality or application, we use it to study beam steering as a particular yet very representative functionality for SDM/RIS-enabled wireless communications [see Fig. 1(a)]. Third, with the help of appropriate figures of merit and subsequent sensitivity analyses, we derive a set of practical design guidelines for the design of efficient programmable MS for beam steering. With this particular case study, we seek to solve questions such as which is the minimum number of unit cells that guarantee a given steering precision over a certain range of angles, or whether it is preferable to put more unit cell states or to make unit cells smaller to improve performance.

The remainder of this paper is organized as follows. In Section \ref{sec:model}, the model for the scalability study is defined. In Section \ref{sec:methodology}, the proposed methodology and the models used for the beam steering case are introduced. The main results of the scalability study are reported in Section \ref{sec:results} and the impact of the incidence and reflection angles on performance are assessed in Section \ref{sec:scanning}. Finally, the main trends and design guidelines arising from this study are discussed Section \ref{sec:discussion}, whereas the paper is concluded in Section \ref{sec:conclusions}.

\section{Scaling Model}
\label{sec:model}
This section outlines the scaling model proposed in this work. The model distinguishes between factors that relate to the MS geometry, Section \ref{sec:Dparams}, as well as the ability to program the MS to match a given application-specific parameter, Section \ref{sec:Pparams}. The model is general, but instantiated here for the case of beam steering.

Figure \ref{Mdl} shows a schematic representation of the system under study. We assume that MSs are deployed to direct reflected rays to a particular user. Each MS has a lateral size of $D_m$ and is composed by a set of reconfigurable unit cells of size $D_u$. The unit cells are driven by a set of controllers, whose function is to choose the states $S_{mn} \in \Sigma,\, \forall m,n$ that will allow to point waves impinging from incidence angles ($\theta_i$, $\varphi_i$) towards a given direction described by ($\theta_r$, $\varphi_r$).
Due to the limited number of states that the unit cells can adopt, i.e. $|\Sigma| = N_{s}$, the theoretically required reflection phase modulation along the MS may not be exactly satisfied, leading to deviations in the reflection direction, i.e. ($\theta_a$, $\varphi_a$) instead of ($\theta_r$, $\varphi_r$), the appearance of side lobes, etc. In what follows, the main parameters are described in more detail.

\begin{figure*}[!t]
\centering
\vspace{-0.5cm}
\includegraphics[width=0.85\textwidth]{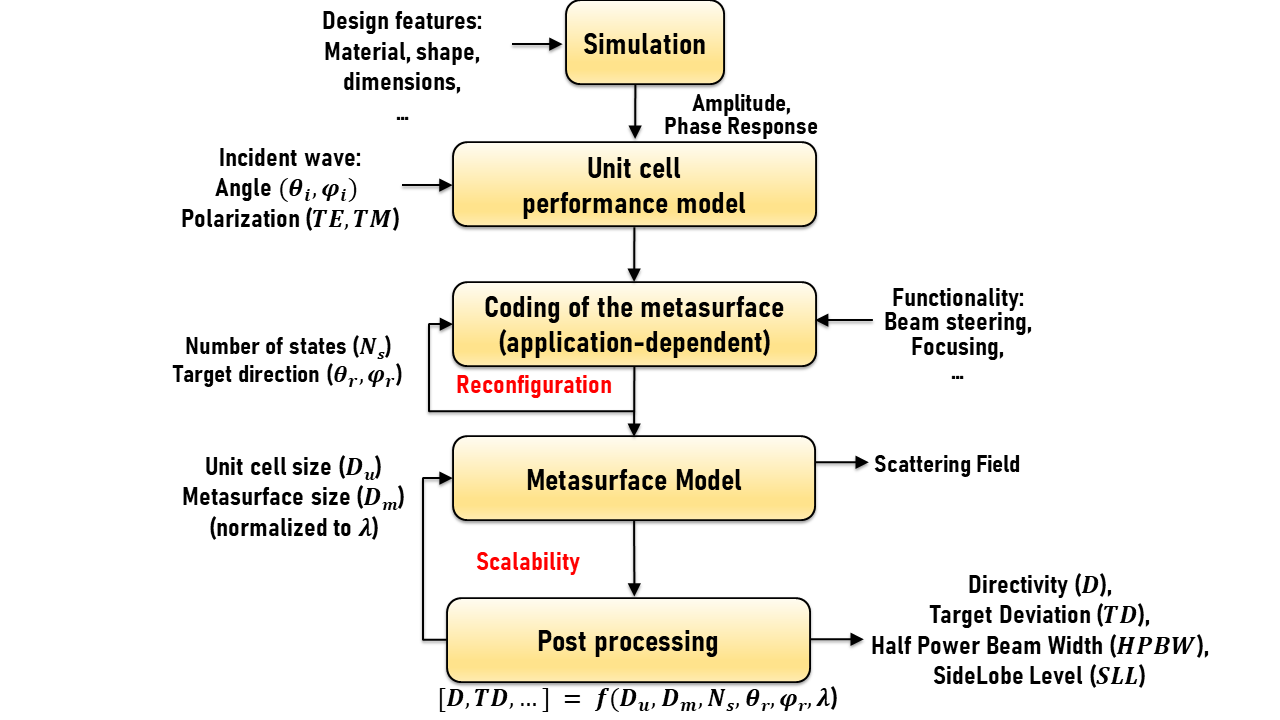}
\vspace{-0.1cm}
\caption{Flowchart of the proposed semi-analytic methodology for scalability analysis. Sections \ref{sec:unit cell} to \ref{sec:metrics} describe each step in detail.}
\label{Mth}
\end{figure*}

\subsection{Dimensional factors}
\label{sec:Dparams}

\noindent \textbf{Size of the unit cell ($D_u$):} The unit cell dimensions commonly depend on the desired frequency regime as they need to be subwavelength. Beyond that, and since the MS is spatially discretized on a unit cell granularity, the size of each unit cell will have an impact on the MS performance. Here, without loss of generality, we assume square unit cells of side $D_u$.

\noindento \textbf{Size of the metasurface ($D_m$):} The size of the MS determines its aperture and ability to coat objects or walls, as well as its cost. Here, we assume that the MS covers a square area with lateral size of $D_m$. With $D_m$ and $D_u$, one can calculate the number of unit cells.

\noindento \textbf{Wavelength ($\lambda$):} From the EM perspective, determining the frequency band of interest is critical to tackle the design of the unit cell. In the case of SDM/RIS-enabled communications, $\lambda$ corresponds to the wavelength in the medium enclosing the MS, typically free space. In our study, instead of adding frequency as another parameter, we express the dimensions normalized to the wavelength in order to give a clear and general vision over the frequency-to-dimensions relationship.

\subsection{Programming parameters}
\label{sec:Pparams}

\noindent \textbf{Number of unit cell states ($N_s$):} Ideally, a programmable MS would have continuous control over the local phase and amplitude of the unit cell responses. However, complexity issues related to the tuning elements and their driving methods often suggest discretizing the amplitude-phase states of the unit cells. The parameter $N_{s}$ that models the number of possible unit cell states is decided at design stage and cannot be modified at runtime. The discretization imposed by the finite number of states will have an impact on the MS performance. Note that, as will be shown in Sections \ref{sec:unit cell} and \ref{sec:Inc}, a pool of available states larger than $N_{s}$ is in generally needed, from which the optimum $N_{s}$ states are chosen for each specific case. This is useful for example for combating the effect of varying incidence angle on the steering performance.


\noindento\textbf{Target direction ($\theta_r$, $\varphi_r$):} As any reflectarray, programmable MSs for beam steering naturally have the direction of reflection as the main input. 
We express the direction using the spherical notation ($\theta_r$, $\varphi_r$) as the position of the intended receiver can be easily expressed in spherical coordinates ($r$, $\theta$, $\varphi$) as well, using the MS as point of reference in the coordinate system as shown in Fig. \ref{Mdl}. Without loss of generality, we assume plane wave incidence and a distant receiver, which allows to define the position of the receiver with $\{\theta_r$, $\varphi_r\}$ only. The model, however, would admit arbitrary wavefront shapes if necessary.

\noindento \textbf{Incidence angle ($\theta_i$, $\varphi_i$):} The unit cell states leading to the desired reflection direction also depend on the angle of incidence. With the assumptions made above, the incidence is fully defined by angles ($\theta_i$, $\varphi_i$) as shown in Fig. \ref{Mdl}. Again, if needed, the model would admit arbitrary wavefront shapes.

We note that, while the number of states is fixed at design time, the incidence angle and target direction will be generally time-variant in SDM/RIS scenarios. For instance, a SDM/RIS designed to add beams coherently at the receiving end will need to adapt the incidence and reflected angle to the positions of the transmitters and receivers.

\section{Methodology}
\label{sec:methodology}
To rigorously calculate the actual reflection phase and amplitude of each discrete state, we consider a single unit cell with periodic Floquet boundary conditions, meaning that an infinite uniform MS comprised of such unit cells is assumed in the simulation. This allows us to perform accurate full wave simulations. When moving to the actual steering MS which is comprised of different unit cells in a supercell configuration, we use the calculated global reflection phase/amplitude states as \emph{local} quantities.  This so-called ``periodic'' approximation is justified by the slowly varying modulation of the MS properties and is frequently used in gradient MS design with excellent results \cite{Tsilipakos:2018aom,Liu:2019}.

To obtain the far field (radiation) pattern of the actual finite-size steering MS, we do not use a full-wave simulation setup, as it can become extremely intensive computationally for large MSs and is thus ill-suited for our scalability analysis where the geometric parameters are scaled by orders of magnitude with a huger number of possible parameter combinations. To bridge this gap, the proposed methodology employs a semi-analytical approach where, as described in detail in what follows, the unit cell response is extracted from physical full-wave simulations while the MS response is calculated analytically using the Huygens' principle.

Figure \ref{Mth} summarizes the action points of the proposed methodology. First, unit cell is designed in a full-wave simulation, then reflection factors are incorporated into the analytic formulation to model the MS. Finally, by processing the scattered field, performance metrics are extracted. Without any compromise on the generalization, the methodology is instantiated to study the case of anomalous reflection for beam steering applications. It can be employed to study practically any wavefront transformation by adopting the corresponding phase gradient and adjusting the selected performance metrics. Sections \ref{sec:unit cell} and \ref{sec:metasurface} describe the unit cell and MS models, respectively. Sections \ref{sec:coding} and \ref{sec:metrics} outline the methods used to derive the optimal coding of the MS for beam steering and the performance metrics that this work considers. Finally, Section \ref{sec:validation} validates the proposed analytical approach and MS coding method.

\begin{figure}[!t]
\vspace{-0.7cm}
\includegraphics[width=0.9\columnwidth]{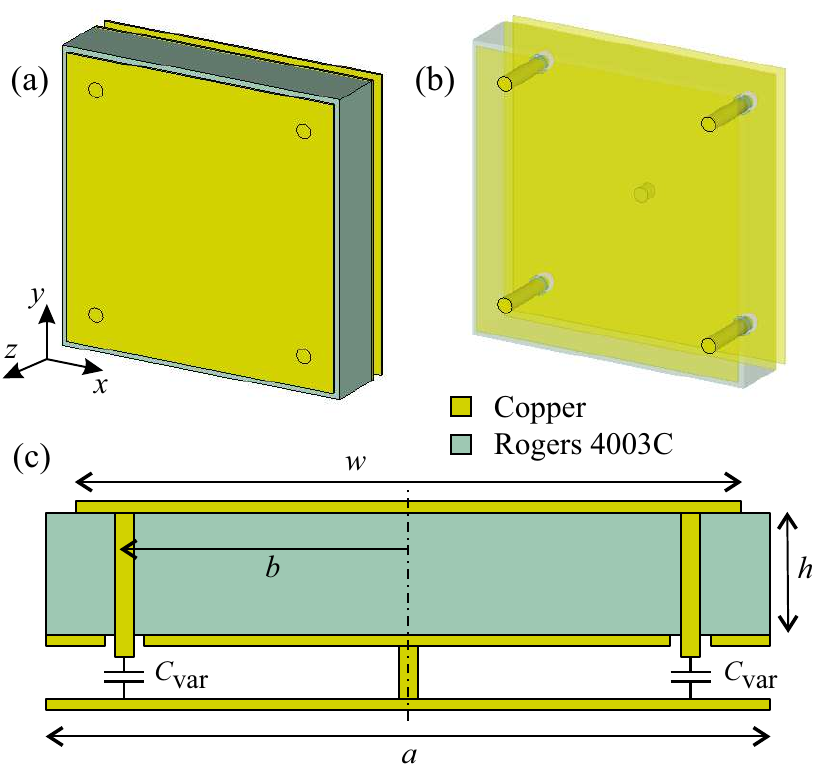}
\vspace{-0.2cm}
\caption{Schematic of unit cell for operation at 26~GHz.  
(a,b)~Bird's eye views indicating the positions of the through vias and the shorting post connecting the chip ground with the metasurface backplane. (c)~Cross-section with annotations of geometric parameters and the varactor capacitances.}
\vspace{-0.2cm}
\label{Cell}
\end{figure}

\subsection{Unit cell performance model}
\label{sec:unit cell}
In this Section, we propose a reconfigurable unit cell for operation in reflection, Fig.~\ref{Cell}. A square unit cell ($a=4$~mm) with a metallic back plane is designed to resonate at $26$~GHz, a band of great interest for 5G applications, and thus provide the necessary $2\pi$ phase delay for implementing wavefront control based on the Huygens' principle. We stress that this physical concept is independent of the adopted physical system and frequency range; for example, a dielectric meta-atom can be used for providing a resonance in the near/far-infrared, or a plasmonic meta-atom for a resonance in the optical regime. 

A square metallic patch ($w=3.92$~mm) is stacked on top of a substrate made of Rogers RO4003C high-frequency board material with permittivity $\epsilon_r=3.38$ and thickness $h=0.203$~mm. The reconfigurability is voltage-controlled and stems from varactor elements properly incorporated in the unit cell, Fig.~\ref{Cell}(c). More specifically, through vias connect the rectangular patch to four varactors residing behind the backplane inside an integrated chip, making it possible to tune the surface impedance of the MS and, thus, the local reflection phase and amplitude. The four vias are positioned in a symmetric fashion near the four corners of the patch, with a distance from the unit cell center along both axes of $b=1.5$~mm, and have a diameter $d=0.1$~mm. The ground of the chip is connected with the MS backplane via a metallic post in the center of the unit cell, Fig.~\ref{Cell}(c). The four varactors are collectively set to the same capacitance value $C_\mathrm{var}$; they are used instead of a single varactor at the center of the unit cell \cite{Liu:2019} in order to enhance the impact of varying capacitance over the surface impedance (induced currents are maximized at the edges of the patch) while retaining an isotropic unit cell (same behavior along both cartesian axes).

\begin{figure}[!ht]
\vspace{-0.4cm}
\includegraphics[width=1\columnwidth]{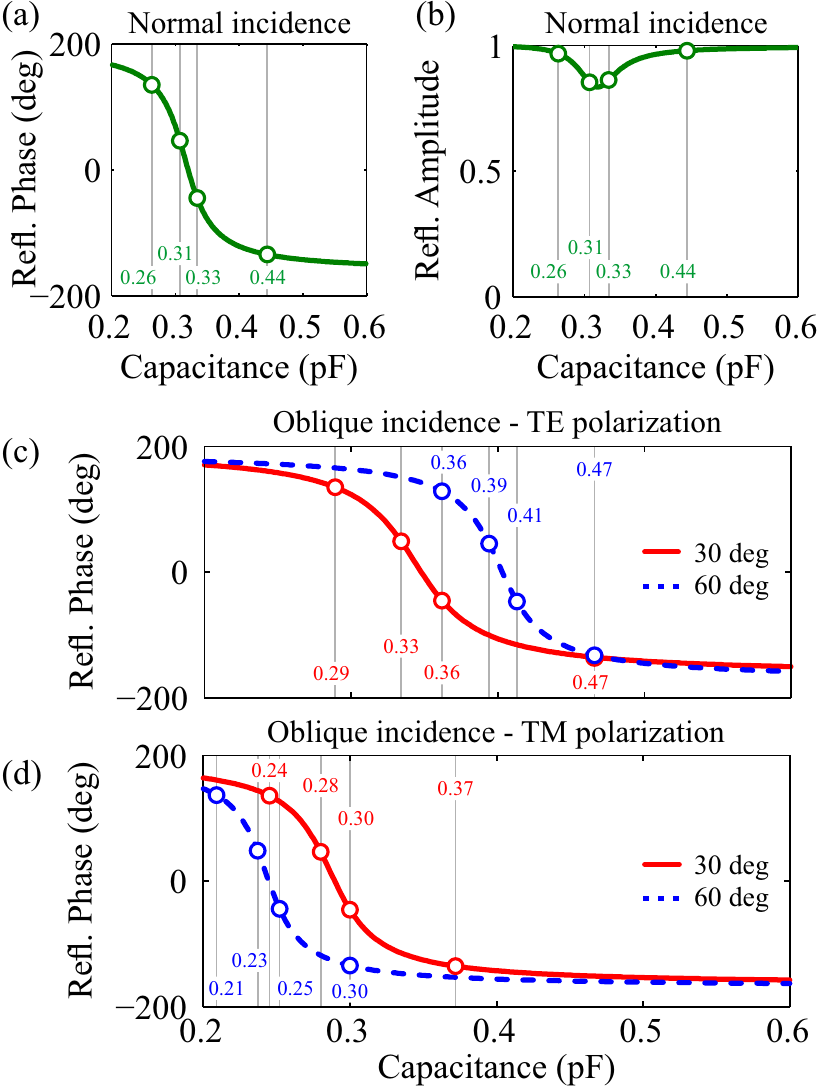} 
\caption{(a) Reflection phase and (b) amplitude for the proposed unit cell under normal incidence as a function of capacitance. The four capacitance values leading to reflection phase $\{135, 45, -45, -135\}$ degrees are marked. (c)~Reflection phase as a function of capacitance for TE polarization and incidence angles 30 and 60 degrees. The capacitance values for the desired four phase states are marked. (d)~Reflection phase as a function of capacitance for TM polarization and incidence angles 30 and 60 degrees. The capacitance values should for the desired four phase states are marked.}
\vspace{-0.1cm}
\label{AP}
\end{figure}

For providing reconfigurable steering performance, we  will combine unit cells of different reflection phase states; e.g. for the case of two-bit coding we use four different states equidistantly spanning the 0--2$\pi$ range, i.e. with values $\{135, 45, -45, -135\}$ degrees. They can be achieved with specific values of the varactor capacitances $C_\mathrm{var}$ by means of an appropriate biasing voltage. In Fig.~\ref{AP} we depict the reflection phase, Fig.~\ref{AP}(a), and reflection amplitude, Fig.~\ref{AP}(b), of the proposed unit cell, as calculated by full-wave simulations of the unit cell for normal incidence. The required reflection phase states are attained for varactor capacitances  $\{0.26, 0.29, 0.31, 0.33\}$~pF. At the same time, the corresponding amplitudes are high and quite uniform; absorption is maximized on resonance and thus it is unavoidable that certain capacitance values that bring the MS resonance closer to $26$~GHz will be associated with smaller reflection amplitudes. The designed phase states can be used to steer a reflected beam towards the desired direction; the performance of this operation will be thoroughly assessed in the following sections. Note that although designed for a specific pitch value $a\equiv D_u$, the proposed unit-cell extent can be scaled and still function around the target frequency of 26~GHz by modifying the required varactor capacitances or, equivalently, the bias voltages. 


Next, we investigate the effect of oblique incidence for both TE and TM polarizations. Specifically, it is expected that the attained reflection phase will depend on the incident angle. This means that the aforementioned capacitance values will provide suboptimal reflection phase as the incidence angle varies. Having at our disposal a different set of four phase states (for the case of two-bit coding) can help in retaining excellent performance for different incidence angles. This is shown in Fig.~\ref{AP}, where the reflection phase as a function of capacitance is depicted for incidence angles of 30 and 60 degrees, for TE [Fig.~\ref{AP}(b)] and TM [Fig.~\ref{AP}(c)] polarization, respectively. By selecting each time the best four out of a total of $16$ available states enables us to retain almost perfect performance for all the cases investigated in Fig.~\ref{AP}.

\subsection{Metasurface coding}
\label{sec:coding}
The direction of reflection can be engineered by an appropriate linear phase gradient \cite{Yu2011a,Tsilipakos:2018aom,Liu:2019}. Assuming that the MS imposes the phase profile $\Phi(x,y)$, we assign the virtual wave vector $\mathbf{k}_\Phi=\nabla\Phi=\partial_x \Phi\,\hat{x} + \partial_y \Phi\,\hat{y}$ ($\partial_x$ and $\partial_y$ denote partial derivatives). The momentum conservation law can be expressed as

\begin{equation}\label{eq:dphi}
\begin{array}{l}
k \sin{\theta_{i}}\cos{\varphi_{i}} + \partial_x\Phi = k \sin{\theta_{r}}\cos{\varphi_{r}}, \\
k \sin{\theta_{i}}\sin{\varphi_{i}} + \partial_y\Phi = k \sin{\theta_{r}}\sin{\varphi_{r}},
\end{array} 
\end{equation}
where $\partial_x\Phi$ and $\partial_y\Phi$ describe the imposed phase gradients in the $x$ and $y$ directions, respectively, and the subscripts $i$ and $r$ denote incident and reflected (scattered) waves, respectively.

\begin{figure}[!ht]
\includegraphics[width=0.32\columnwidth]{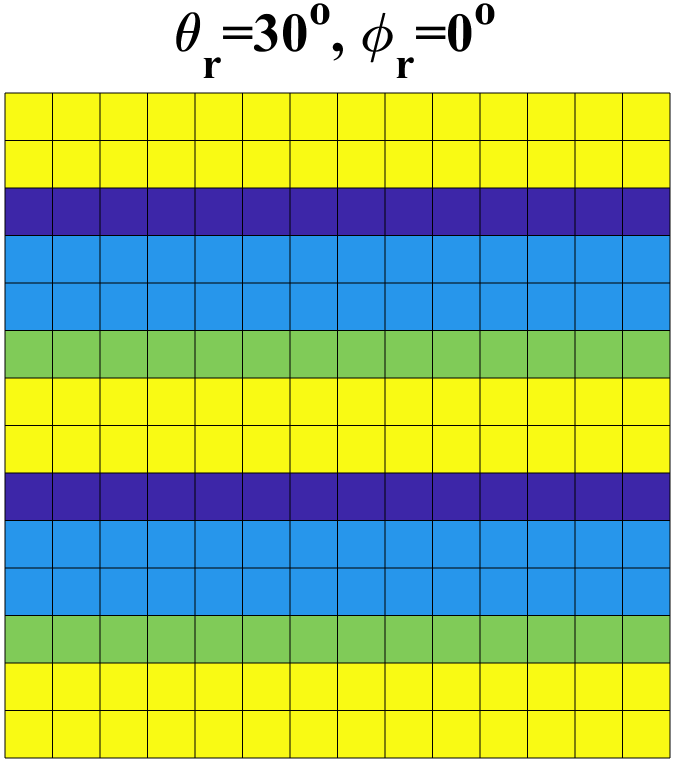} 
\includegraphics[width=0.32\columnwidth]{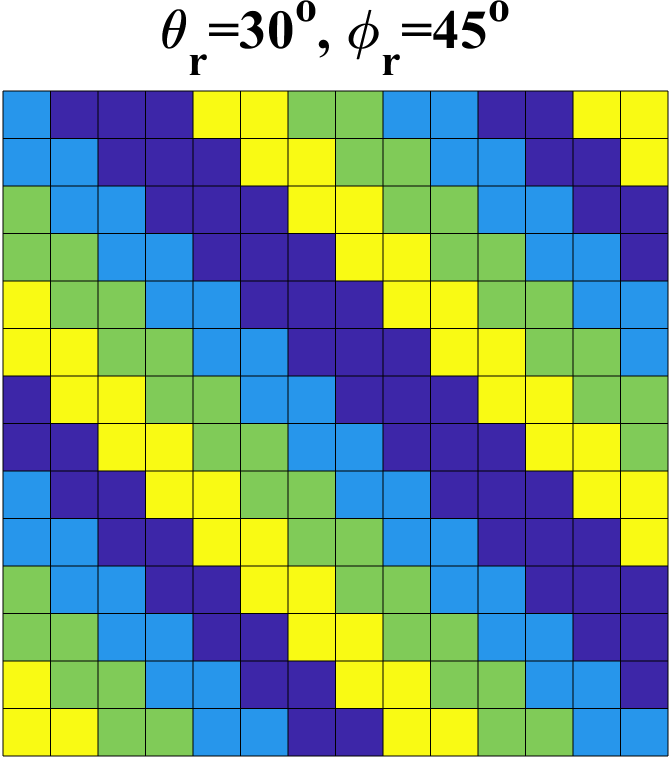} 
\includegraphics[width=0.32\columnwidth]{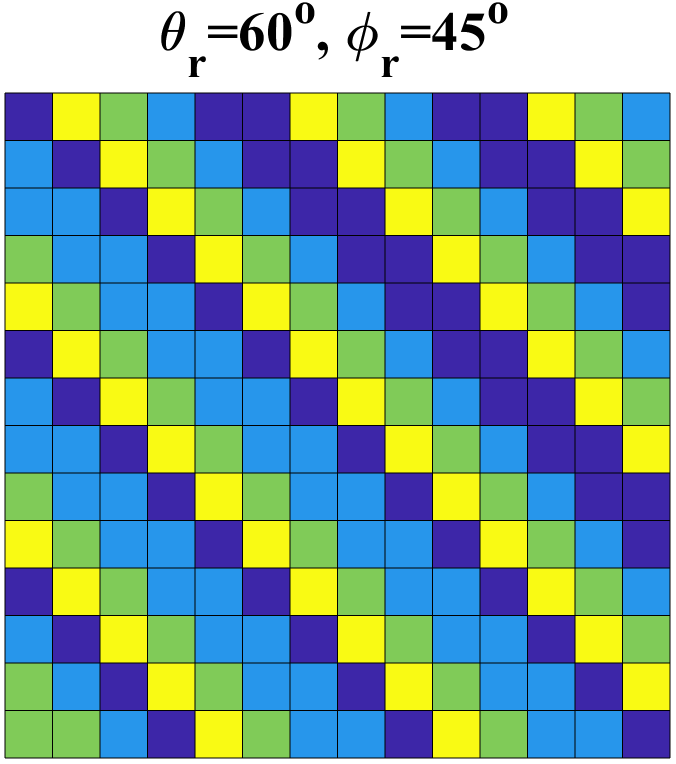} 
\caption{Coding of a $15\times15$ metasurface with $N_{s} = 4$ for different desired reflection angles assuming normal incidence. Each color represents a different state (blue: 00, yellow: 01, cyan: 10, green: 11) with equispaced reflection coefficient phases.}
\label{MC}
\end{figure}

To simulate the MS and perform the scalability analysis, the applied coding should yield the best possible performance across different physical scales. Our approach, instead of relying on fixed super-cell or meta-atom structures \cite{Zhou2018}, calculates the phase gradient at the unit cell granularity and adapts the unit cells states accordingly. Therefore, we fix the unit cell size ($d_x=d_y=D_u$) and then obtain the phase required at the $mn$-th unit cell. Assuming air as the host medium the required phase reads
\begin{equation}
\Phi_{mn}=\frac{2\pi D_u (m\cos\varphi_r \sin\theta_r+n\sin\varphi_r \sin\theta_r)}{\lambda_0}  
\label{eq:phimn}
\end{equation}
Subsequently, based on the number of unit cell states $N_{s}$ and the phase gradient profile, the nearest available state will be mapped to the unit cell. Note that to adapt to the digital logic of the control devices, the number of states is associated to the number of bits $N_b$ used to encode the states through $N_{s} = 2^{N_b}$. Depending on $N_b$, the phase states are separated by $\pi/2^{N_b-1}$ in the $2\pi$ range. For example, a 2-bit coding MS possesses 4 pahse states (``00'', ``01'', ``10'' and ``11'') which are $0$, $\pi/2$, $\pi$, $3\pi/2$. Note that a constant phase offset for all states would not change the performance, it is the phase difference between states that is important. To illustrate the output of the coding process and the impact of the deflection angles $(\theta,\varphi)$ on the required phase gradients in the $x$ and $y$ directions, Fig.~\ref{MC} depicts the MS phase profile for different pairs of target angles assuming normal incidence. 

\subsection{Metasurface model}
\label{sec:metasurface}
Following the Huygens principle in the far-field limit, the MS cells can be accurately modeled as a collection of sources of secondary radiation. For linearly polarized incidence, the scattered field can be expressed as \cite{Yang2016a}
\begin{equation}
\begin{split}
E(\theta, \varphi) = \sum_{m=1}^{M} \sum_{n=1}^{N}A_{mn}e^{j\alpha_{mn}} f_{mn}(\theta_{mn}, \varphi_{mn})\\
\Gamma_{mn}e^{j\Phi_{mn}}f_{mn}(\theta, \varphi)
e^{jk_0\zeta_{mn}(\theta, \varphi)}
\end{split}
\label{eq1}
\end{equation}
where $\varphi$ and $\theta$ are the azimuth and elevation angles, $A_{mn}$ and $\alpha_{mn}$ are the amplitude and phase of the wave incident on the $mn$-th unit cell, $\Gamma_{mn}$ and $\Phi_{mn}$ are amplitude and phase reflection coefficient for the $mn$-th unit cell, and $f_{mn}$ denotes the scattering pattern of the $mn$-th unit cell, which, according to reciprocity, is identical for scattering toward the $(\theta, \varphi)$ direction and the interception of incoming waves from the $(\theta_{mn}, \varphi_{mn})$ direction; here we assume $f_{mn}(\theta, \varphi)=\cos(\theta)$ which describes real-world dipolar scatterers. Finally, $\zeta_{mn}(\theta, \varphi)$ is the relative phase shift of the unit cells with respect to the radiation pattern coordinates, given by
\begin{equation}
\zeta_{mn}(\theta, \varphi) = D_u\sin{\theta}[(m-\tfrac{1}{2})\cos{\varphi}+(n-\tfrac{1}{2}) \sin{\varphi}]
\end{equation}

In summary, after evaluating the phase required at each unit cell using Eq. \eqref{eq:phimn} and performing the nearest neighbour mapping to the available unit cell states, the amplitude and phases from the unit cell performance models are introduced in Eq. \eqref{eq1} through $\Gamma_{mn}$ and $\Phi_{mn}$ to obtain the far-field pattern of the MS.



\subsection{Performance metrics}
\label{sec:metrics}
The far field pattern obtained in the previous step is post-processed to obtain a set of performance metrics relevant to beam steering. We detail them next.

\noindento \textbf{Directivity ($D(\theta, \varphi)$):} A fundamental antenna parameter quantifying concentration of energy at a given direction with respect to isotropic scattering, calculated as
\begin{equation}
D(\theta, \varphi)=\frac{4\pi U(\theta, \varphi)}{\int_{0}^{2\pi}\int_{0}^{\pi} U(\theta, \varphi) \sin\theta d\theta d\varphi},
  \label{direc}
\end{equation}
where $U(\theta, \varphi)\propto |E(\theta, \varphi)|^2$ is the radiation intensity scattered towards a given direction, and the denominator corresponds to the total scattered power. For a fully reflective MS, the elevation angle $\theta$ is limited to $[0,\pi/2]$ while the maximum directivity is limited to $4\pi A /\lambda^2$, where $A$ is the MS aperture area. In the results section, we evaluate the directivity in relevant angles such as the target reflection angle $(\theta_r, \varphi_r)$ and the actual reflection angle $(\theta_a, \varphi_a)$ (see Fig. \ref{Mdl}).

\noindento  \textbf{Target deviation ($TD$):} It is measured in degrees and quantifies the difference between the target ($\theta_r, \varphi_r$) and the actual ($\theta_a,\varphi_a$) reflected angle due to inaccuracies in the phase profile. It is  calculated as
\begin{equation}
TD=\sqrt{(\theta_r - \theta_a)^2+(\varphi_r - \varphi_a)^2}.
\end{equation}

\noindento  \textbf{Side-lobe level ($SLL$):} In addition to the main beam, a set of minor reflected beams may arise due to the phase profile of the MS and, especially, its finite aperture. The $SLL$ is defined as the ratio (in dB) of the directivity of the side-lobe nearest to the main lobe. A low $SLL$ is preferable to minimize scattering of energy in unwanted directions. For fully reflective MS, a best case of $SLL\approx -13.5$~dB is anticipated.

\noindento \textbf{Half power beam width ($HPBW$):} The waist of the main reflected beam defines the resolution of steering. The $HPBW$, measured in degrees, is calculated as the square root of the solid angle at the $-3$~dB of a lobe maximum. Low values suggest very accurate localization and tracking, whereas high values suggest diffuse scattering or higher angular coverage.

\begin{figure}[!t]
\centering
\includegraphics[width=\columnwidth]{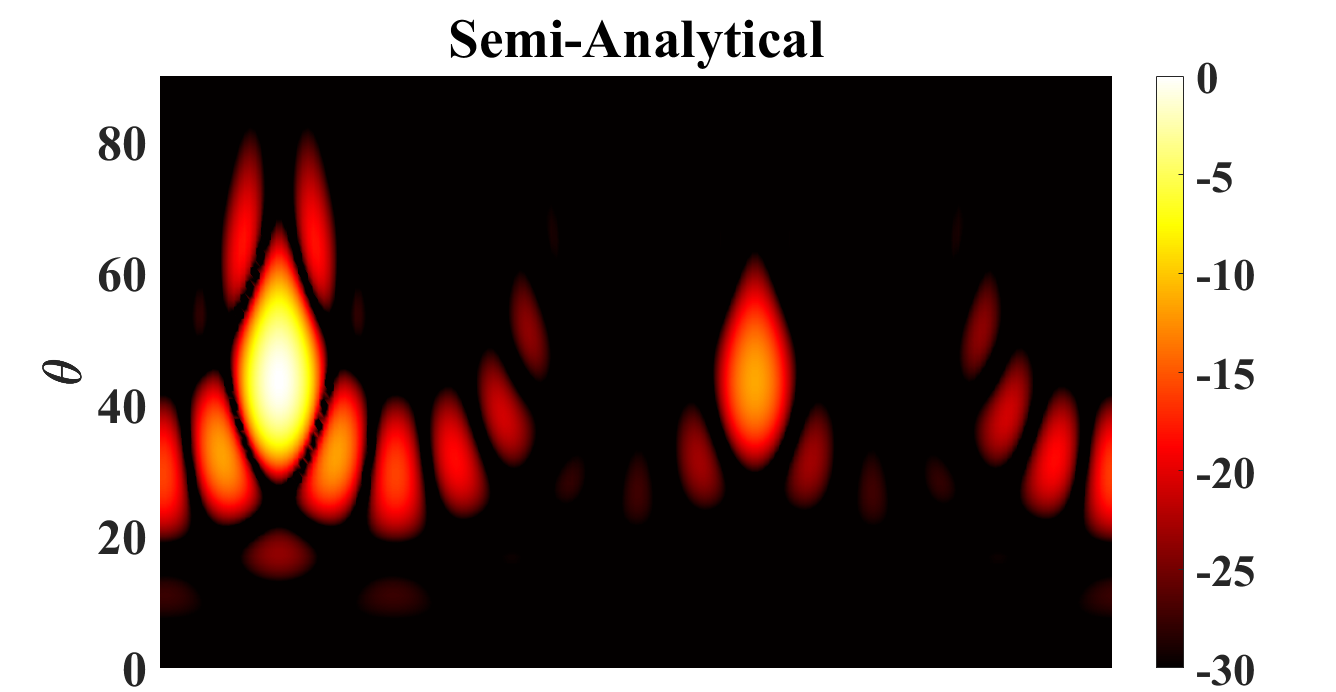}
\includegraphics[width=\columnwidth]{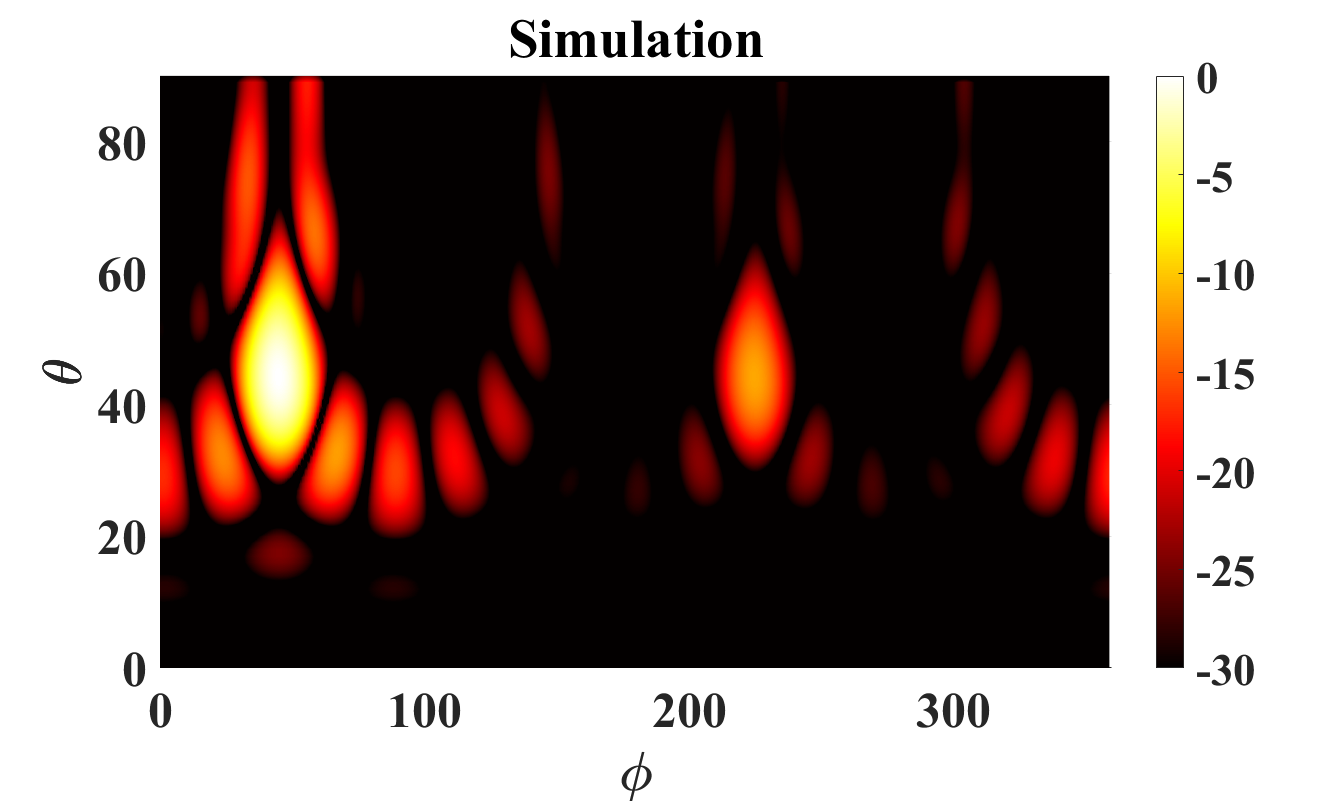}
\vspace{-0.4cm}
\caption{Normalized power radiation (E-Field, dB) of the programmable metasurface while targeting $\theta_r=\varphi_r=\pi/4$, calculated with our method (top) and full-wave simulation (bottom). Excellent agreement is observed.}
\vspace{-0.2cm}
\label{S}
\end{figure}

\begin{figure*}[!t]
\centering
\includegraphics[width=.67\columnwidth]{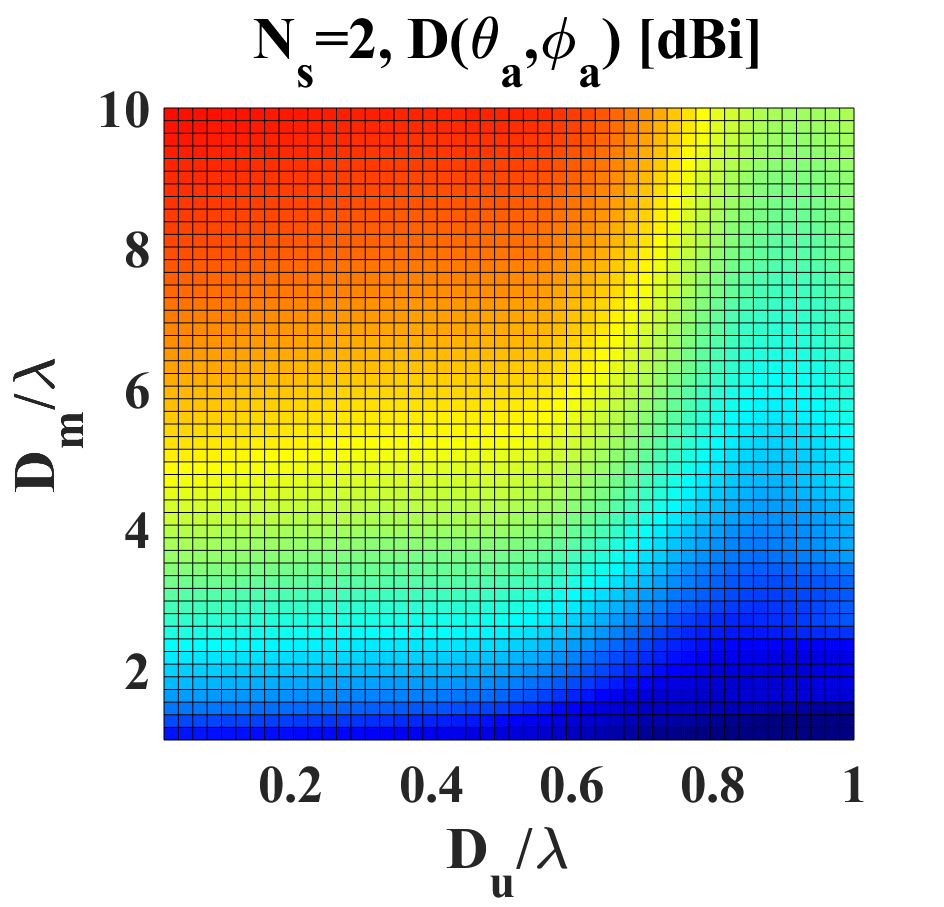}
\includegraphics[width=.67\columnwidth]{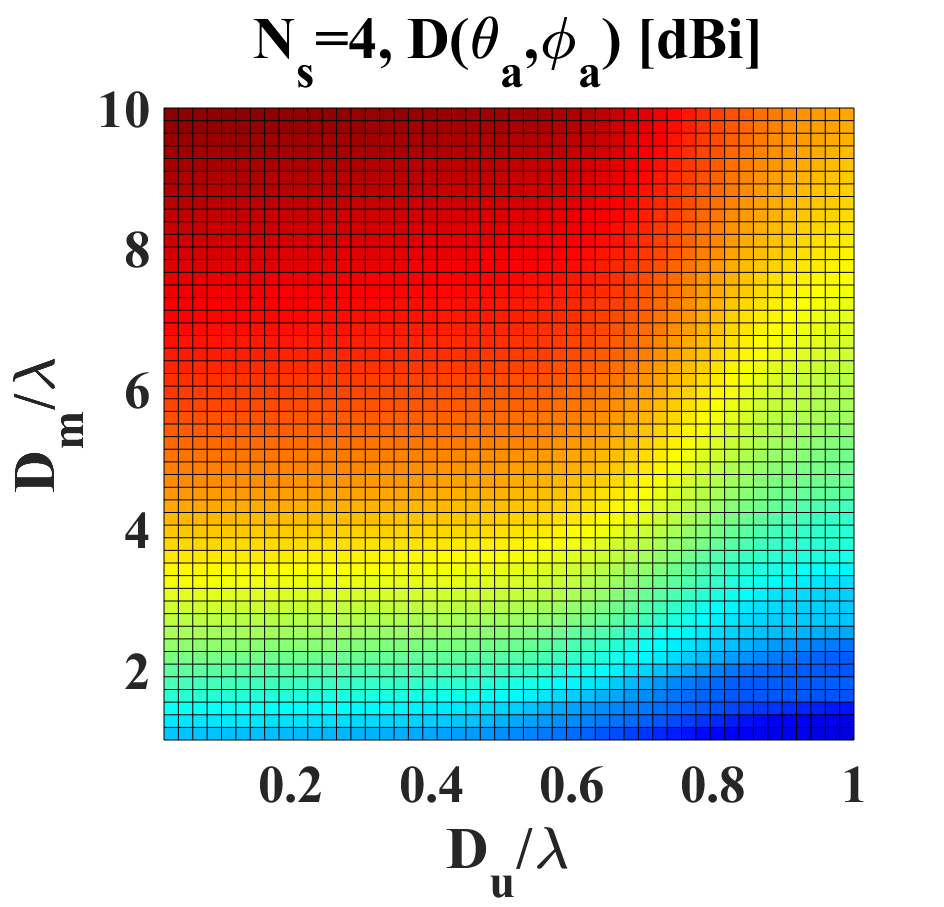}
\includegraphics[width=.67\columnwidth]{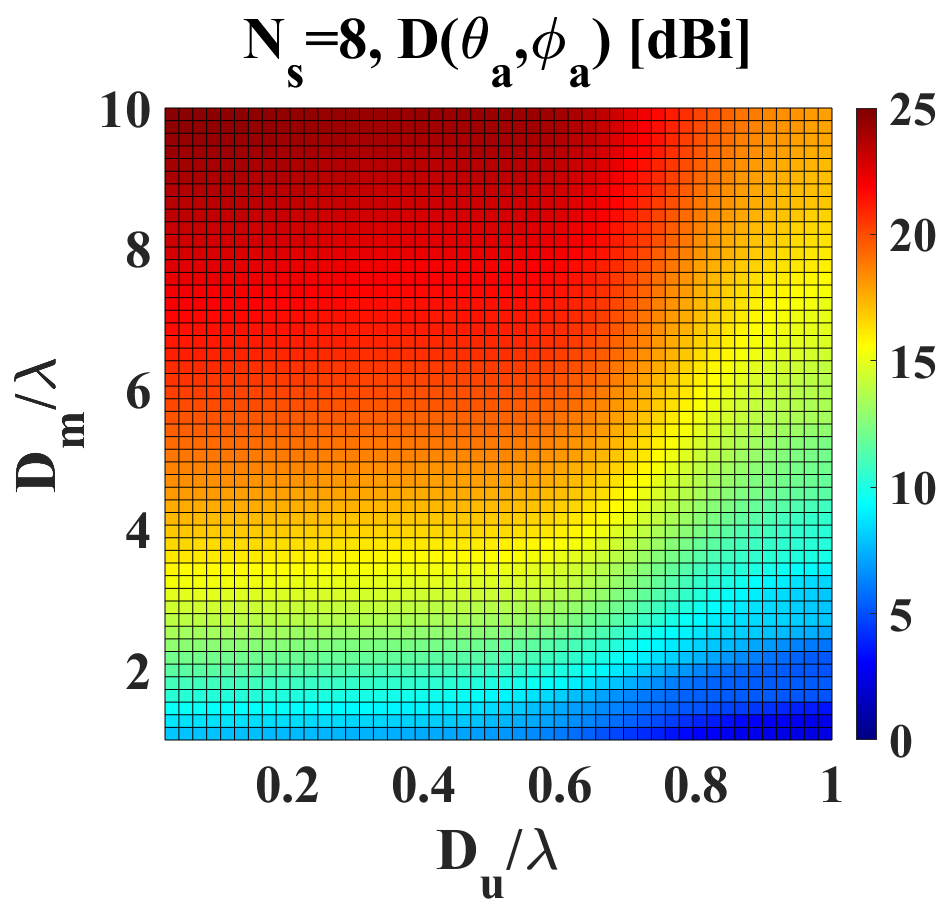}
\vspace{-0.3cm}
\caption{Directivity at the direction of maximum radiation $D(\theta_a,\varphi_a)$ for $\varphi_r=\theta_r=\pi/4$ as a function of the dimensional parameters for 1-bit, 2-bit and 3-bit coding. The color bar is common to all figures.}
\vspace{-0.3cm}
\label{dir}
\end{figure*}

\subsection{Validation}
\label{sec:validation}
The accuracy of the proposed semi-analytical method is verified through a comparison with full-wave simulations by assuming a MS with dimensional parameters $D_u=\lambda/3$ and $D_m=5\lambda$ and a desired reflection angle $\theta_r=\varphi_r=\pi/4$ under normal plane-wave incidence. 
As shown in Fig. \ref{S} the semi-analytical method is in excellent agreement with the full wave simulation. At the same time it is considerably faster and thus perfectly suited to the following scalability study. 



\section{Performance Scalability}
\label{sec:results}
\begin{figure*}[!t]
\includegraphics[width=.67\columnwidth]{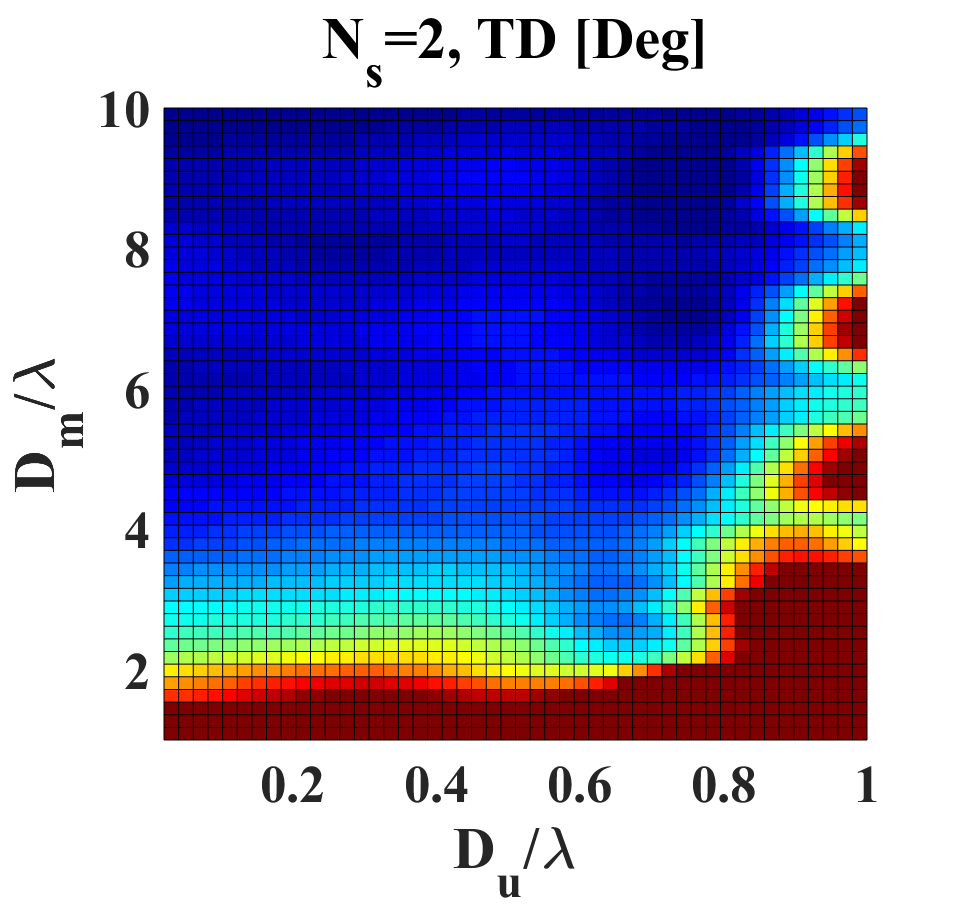}
\includegraphics[width=.67\columnwidth]{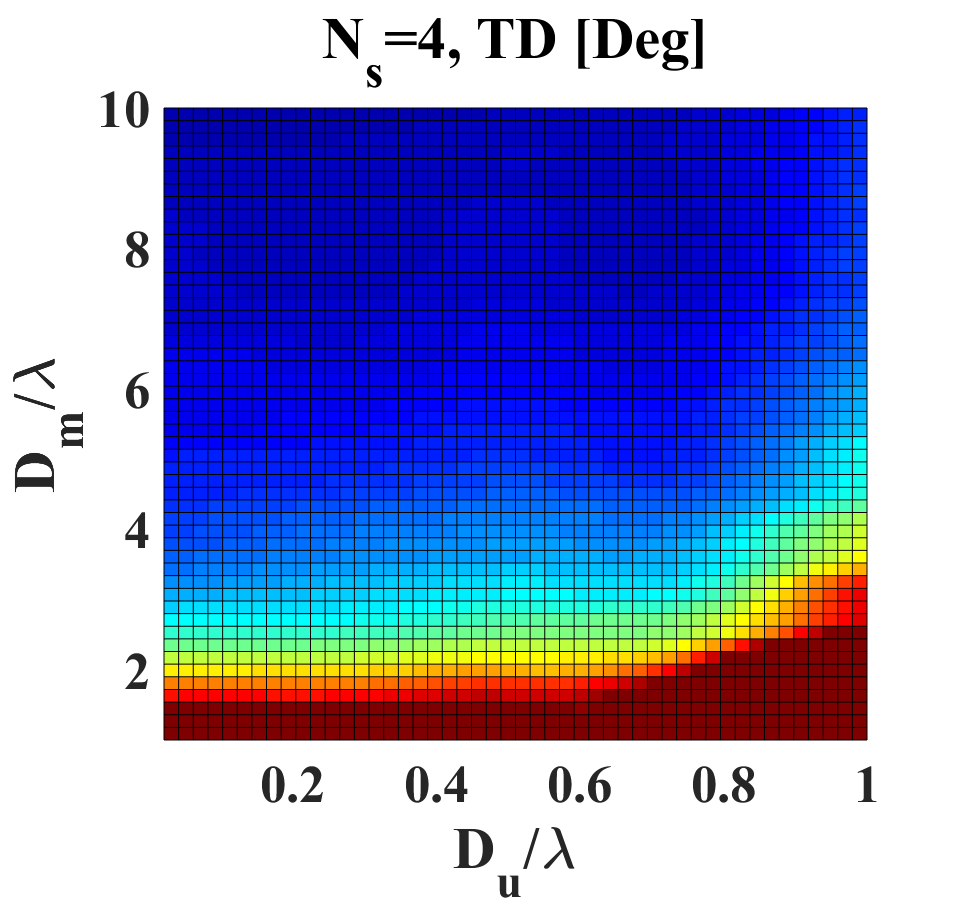}
\includegraphics[width=.68\columnwidth]{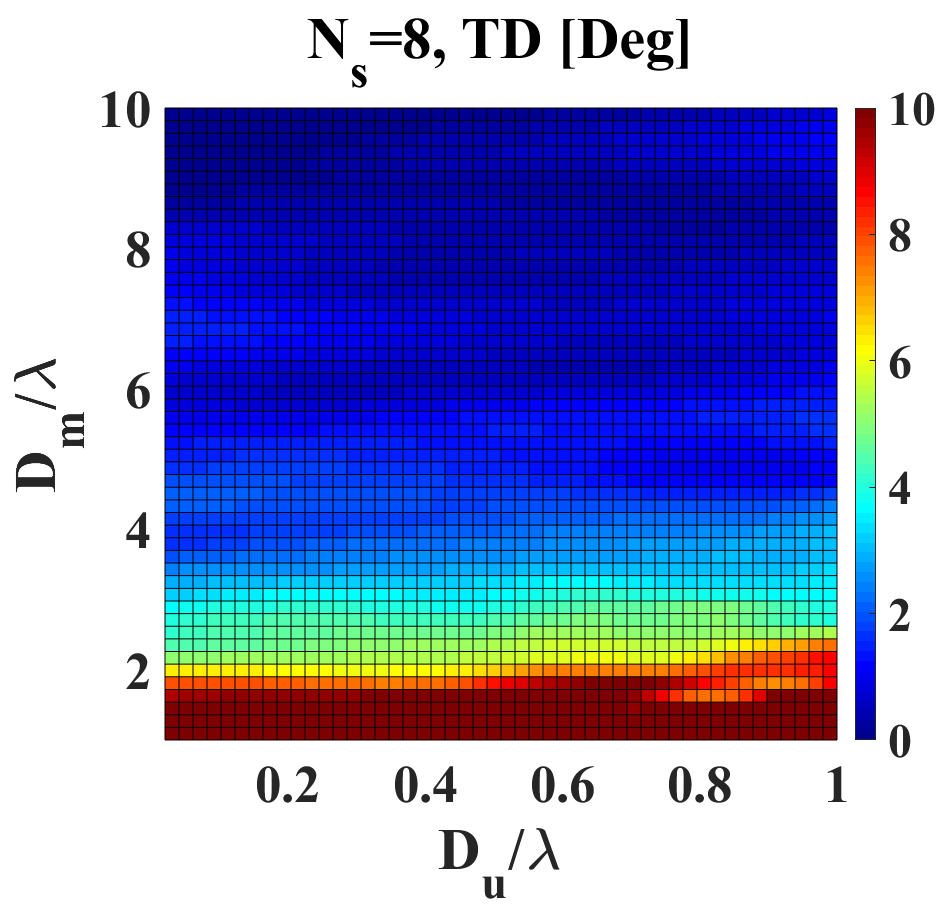}
\vspace{-0.6cm}
\caption{Target Deviation (TD) as a function of dimensional parameters for 1-bit, 2-bit, and 3-bit programmable metasurfaces targeting $\varphi_r=\theta_r=\pi/4$. The color bar is common to all figures.}
\vspace{-0.2cm}
\label{TD}
\end{figure*}

The evaluation of a beam steering system relies on multiple metrics. Here, we obtain the directivity $D$, side-lobe level $SLL$, half power beam width $HPBW$ and target deviation $TD$ as functions of the unit cell size $D_u$, MS size $D_m$, and number of states $N_s$. The parameters are swept by at least an order of magnitude by the definition of scalability analysis. Evidently, some parameter combinations and regions will be unfeasible or \emph{de facto} unacceptable, by virtue of reflect-array principles; nevertheless, this helps to better identify the frontier between relevant and irrelevant design spaces, and highlights the shortcomings of the latter to a broader audience.

To present comprehensive results, we normalize the dimensions to the incident wave wavelength ($\lambda$). This way, the reasoning is applicable to any frequency as long as the scaled unit cell is redesigned to offer the required amplitude-phase response\footnote{Note that the unit-cell phase shifts needed for beam steering have been demonstrated across the spectrum \cite{Munk, Glybovski2016, 6230714}.}. Also, the reported results are for particular target angle $\varphi_r=\theta_r=\pi/4$ and normal incidence. 
The effect of the incidence and target angles on the performance of the MS is discussed later in Section \ref{sec:scanning}.

\subsection{Directivity}
We first assess the directivity in the direction of maximum radiation $(\theta_a, \varphi_a)$ as a function of the three input parameters $D_u$, $D_m$, and $N_s$. 
Figure \ref{dir} shows how the directivity scales with respect to $D_u/\lambda$ and $D_m/\lambda$ for three representative values of $N_{s}$ corresponding to 1-bit, 2-bit, and 3-bit coding. It is observed that the directivity increases with the MS size. For instance, for $N_s = 4$, we see a consistent increase of 15 dB when moving from $D_{m} = \lambda$ to $D_{m} = 3\lambda$. The improvement diminishes from there, yet an additional 10 dB can be achieved when moving from $D_{m} = 3\lambda$ to $D_{m} = 10\lambda$. The impact of the unit cell size is only appreciable above $D_u=\lambda/2$. Reducing the size further does not improve the directivity of the MS, therefore discouraging the use of small unit cells due to the associated raise of the fabrication complexity and cost.


The impact of the number of states is especially noticeable as we move from $N_s=2$ to $N_s=4$, with a general improvement of $\sim$3 dB. The main reason behind this behavior is that, for $N_s=2$, the reflected wave is split into two identical lobes directed to two symmetrical angles and, therefore, half of the power is lost. This behavior disappears when introducing the second bit of coding, which explains the 3 dB difference. Adding a more states beyond $N_{s} = 4$ bit does not have a significant impact.


\subsection{Target Deviation}
Figure \ref{TD} shows the scaling trends of the $TD$, which we generally aim to minimize in order to achieve high steering precision. Here, we consider 10 degrees to be the maximum admissible deviation, although we will see that such interpretation will depend on the beam width as well.

\begin{figure*}[!t]
\includegraphics[width=.67\columnwidth]{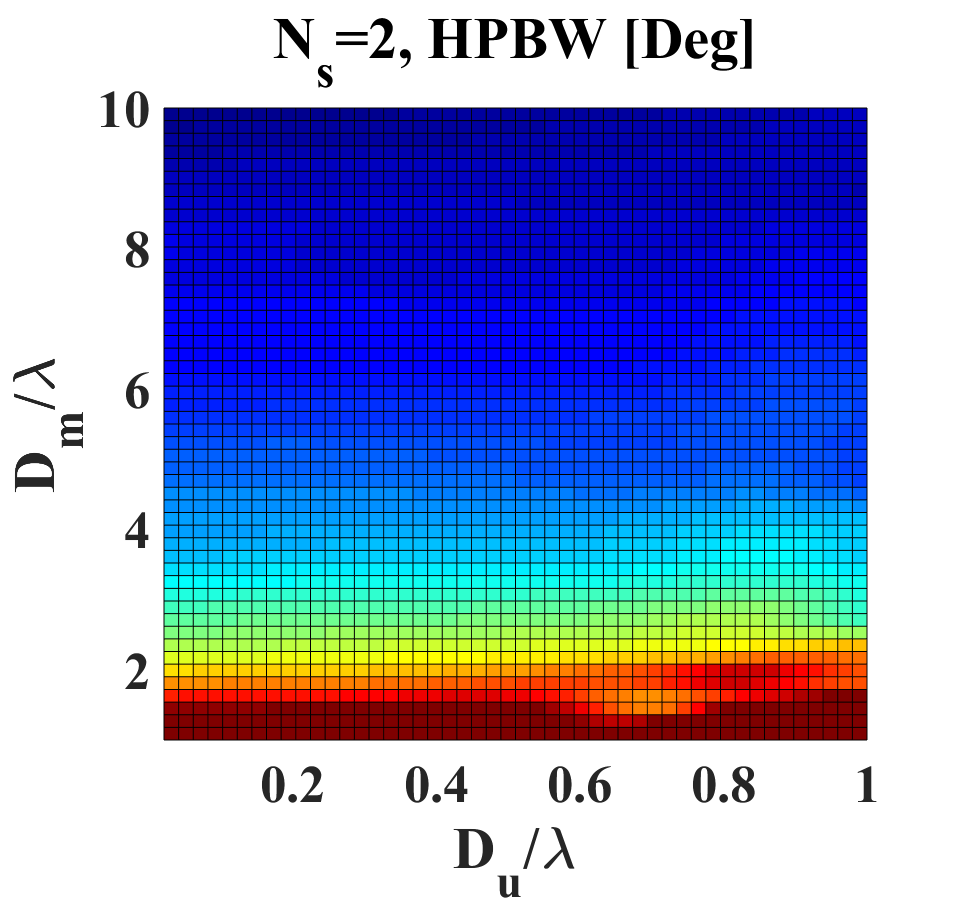}
\includegraphics[width=.67\columnwidth]{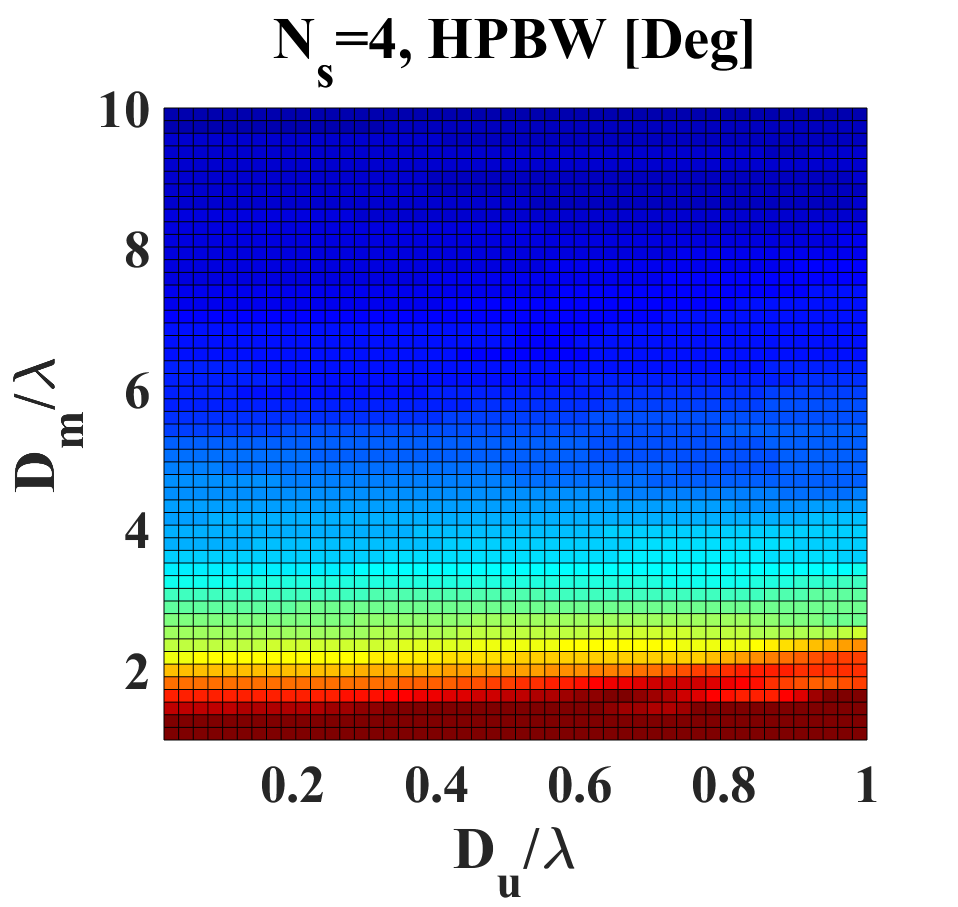}
\includegraphics[width=.67\columnwidth]{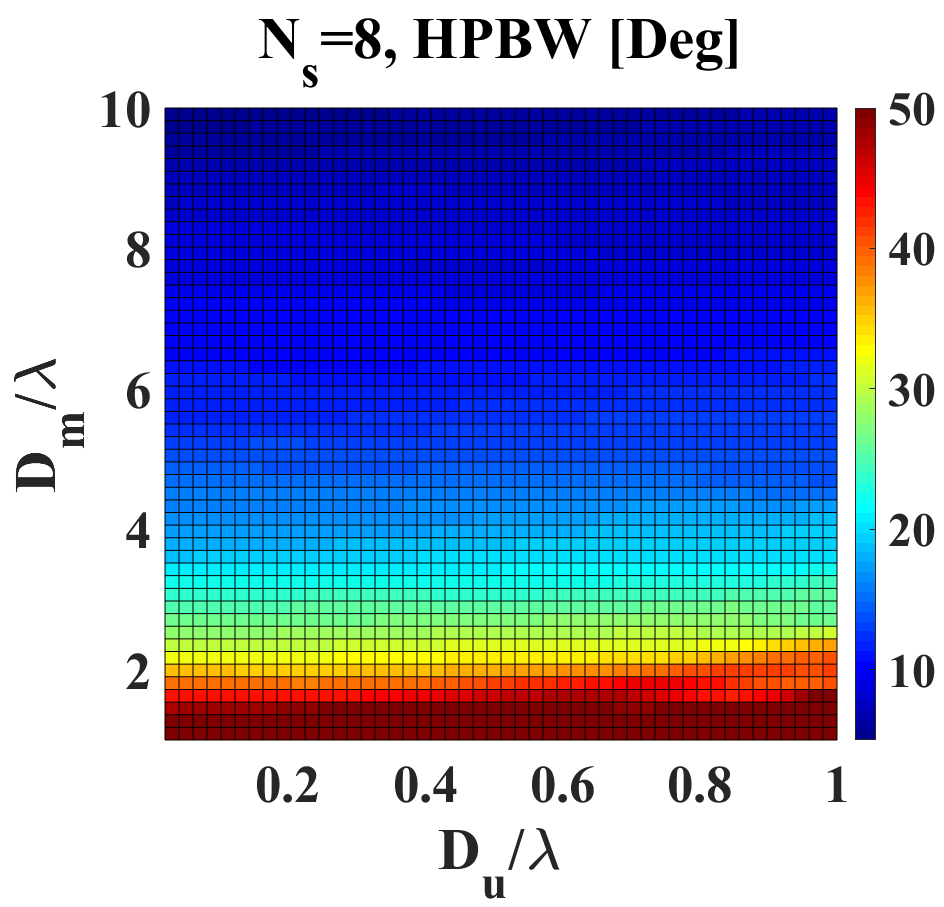}
\vspace{-0.3cm}
\caption{Half Power Beam Width (HPBW) as a function of dimensional parameters for 1-bit, 2-bit, and 3-bit programmable metasurfaces targeting $\varphi_r=\theta_r=\pi/4$. The color bar is common to all figures.}
\vspace{-0.3cm}
\label{HPBW}
\end{figure*}

The results of \ref{TD} demonstrate that $TD$ depends greatly on all the evaluated scaling factors. Downscaling the unit cells diminishes the target deviation of the MS because this implies that the MS is programmed at a finer spatial resolution. However, as in the case of directivity, we observe diminishing results as we reach values around $D_u = \lambda/3$. The impact of the phase quantization error, this is, when increasing the number of states, is also similar than in the directivity case: the improvement is appreciable as we move from $N_s = 2$ to $N_s = 4$, but marginal beyond that. Finally, we note that the impact of the metasurface size $D_m$ is significant only for MS with relatively large unit cells. This implies that one can achieve reasonable steering precision with small MSs as long as the unit cells are also small. 

\begin{figure*}[!t]
\includegraphics[width=.67\columnwidth]{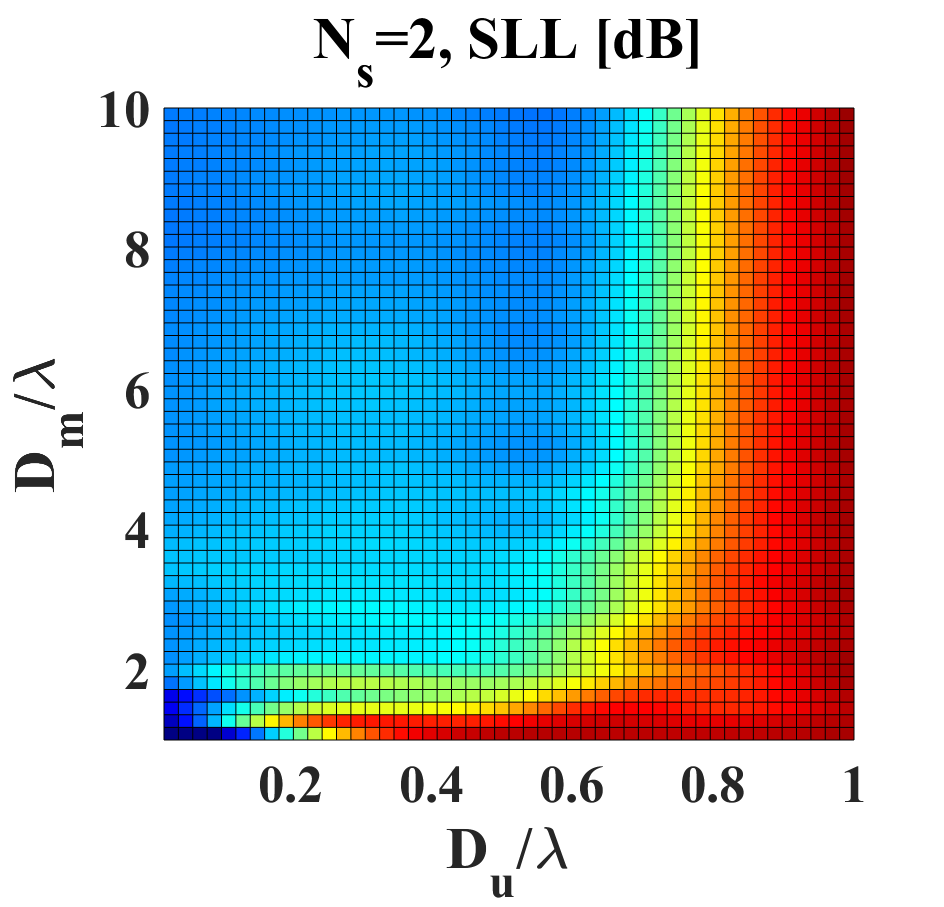}
\includegraphics[width=.67\columnwidth]{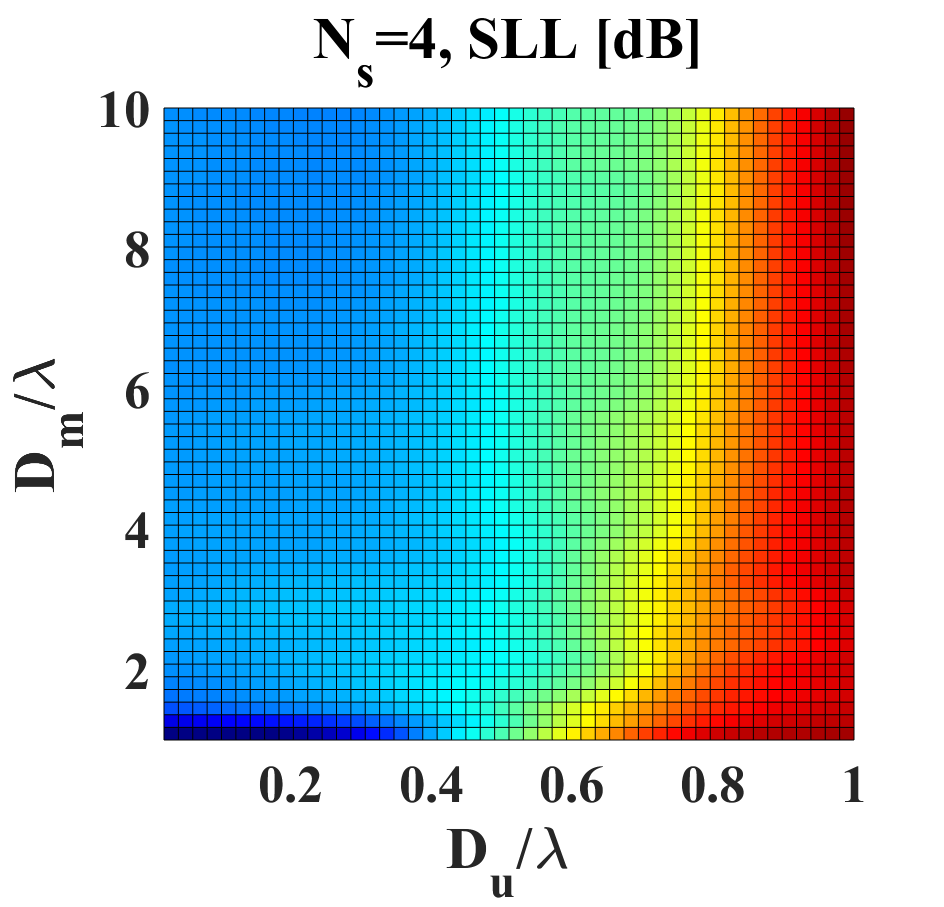}
\includegraphics[width=.67\columnwidth]{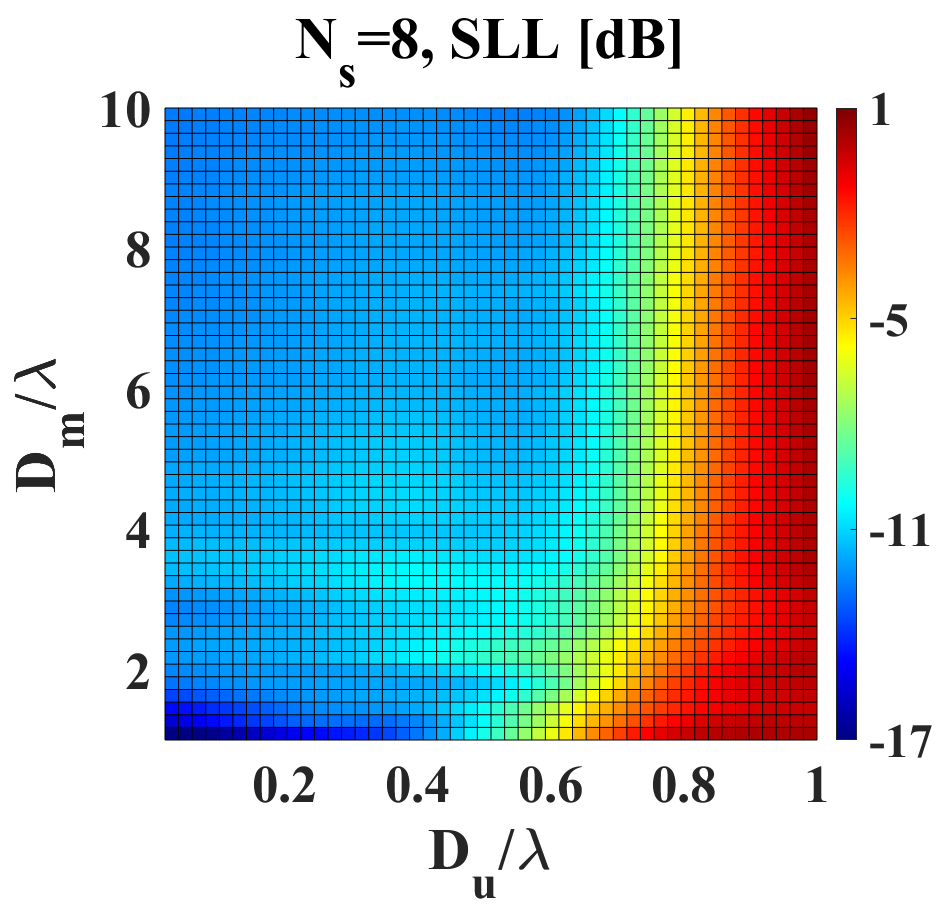}
\vspace{-0.3cm}
\caption{Side-Lobe Level (SLL) as a function of dimensional parameters for 1-bit, 2-bit, and 3-bit programmable metasurfaces targeting  $\varphi_r=\theta_r=\pi/4$. Side-Lobe Level is normalized to the maximum across all MSs. The color bar is common to all figures.}
\vspace{-0.3cm}
\label{SLL}
\end{figure*}
 
%

\subsection{Half Power Beam Width}
The spatial resolution of a steering MS is generally inversely proportional to the $HPBW$, which we aim to reduce. Figure \ref{HPBW} shows how the $HPBW$ is mainly affected by the MS size. This is because the aperture of the device is effectively increased. The improvement is very clear for $D_m < 4\lambda$, to the point that values below 15\textsuperscript{o} are consistently achieved for $D_m \geq 6\lambda$. For a MS of $10\lambda \times10\lambda$, the $HPBW$ is reduced down to around 5\textsuperscript{o}. On the other hand, the impact of the unit cell size and number of states is negligible in this case. 


\subsection{Side Lobe Level}
The evaluation of the $SLL$ is a good first-order estimation of the power that may be off-target and interfere with nearby communications. Figure \ref{SLL} shows the scaling tendencies of $SLL$. Remind that $N_s=2$ is a particular case where the scattered field is split into two identical beams, which would lead to $SLL=0$~dB throughout the design space. Therefore, for this case, we calculate the $SLL$ with respect to the third lobe. For $N_s=4$ and $N_s=8$, the $SLL$ is evaluated as usual.

Figure \ref{SLL} essentially proves that the unit cell size is the main determinant of $SLL$. We can clearly observe how $D_u = \lambda/2$ marks a frontier between a region of good performance in terms of $SLL$ with values below -12 dB from a design space with $SLL$ in excess of -9 dB. It is also worth remarking that, unlike the rest of metrics, the $SLL$ keeps improving as we introduce a third bit of coding ($N_s = 8$). This reinforces the intuition that the $SLL$ is mainly affected by errors in the discretization and quantization of the space-phase. We finally note that, although the MS size does not have a significant influence on this metric, we could compensate the existence of large unit cells with enough unit cell states in large metasurfaces.


\section{Impact of Input/Output Angles on Performance}
\label{sec:scanning}
In this section we will investigate the impact of reflection direction on the steering performance metrics for MS with (i) variable aperture and cell size, but ideal unit cell response across all angles in Section \ref{sec:sensitivity}; and (ii) using a realistic (physical) implementation for the unit cells in Section \ref{sec:Inc}. This way, we differentiate between the performance degradation caused by the MS at large or by individual unit cells. Exploiting the rotational symmetry of the structure and the inherent reciprocity of the EM problem, only a subset of all combinations of incidence (input) and reflection (output) directions needs to be analyzed. Moreover, as highlighted in the previous section, four phase states are sufficient for the basic steering functionality so will limit our simulations to this case and briefly comment on the higher-state cases.

\subsection{Impact on Metasurfaces with Ideal Unit Cells}
\label{sec:sensitivity}
\begin{figure}[!t]
    \includegraphics[width=0.49\columnwidth]{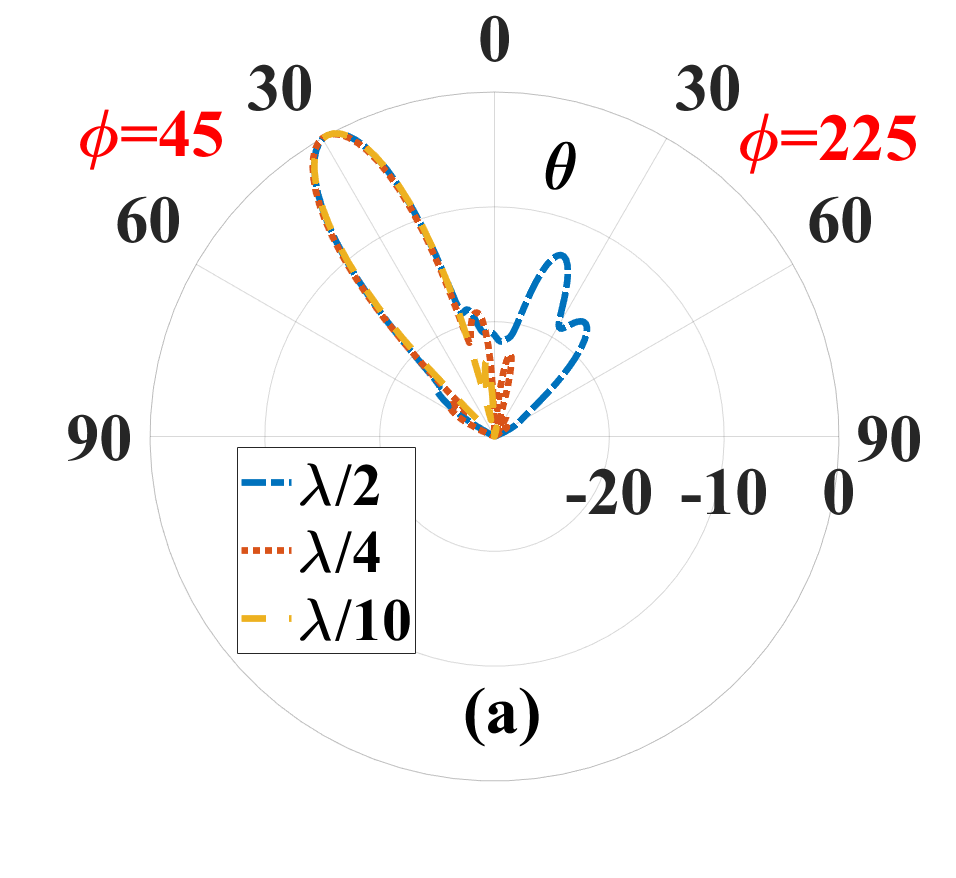}
    \includegraphics[width=0.49\columnwidth]{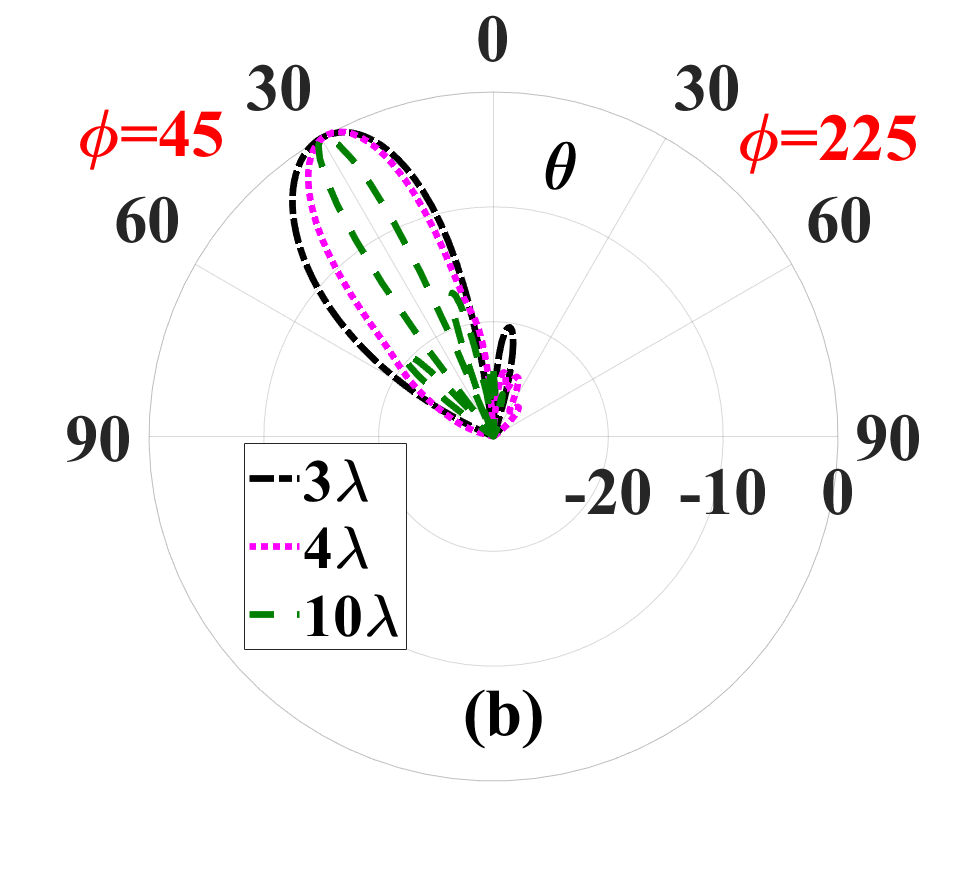}
    \includegraphics[width=0.49\columnwidth]{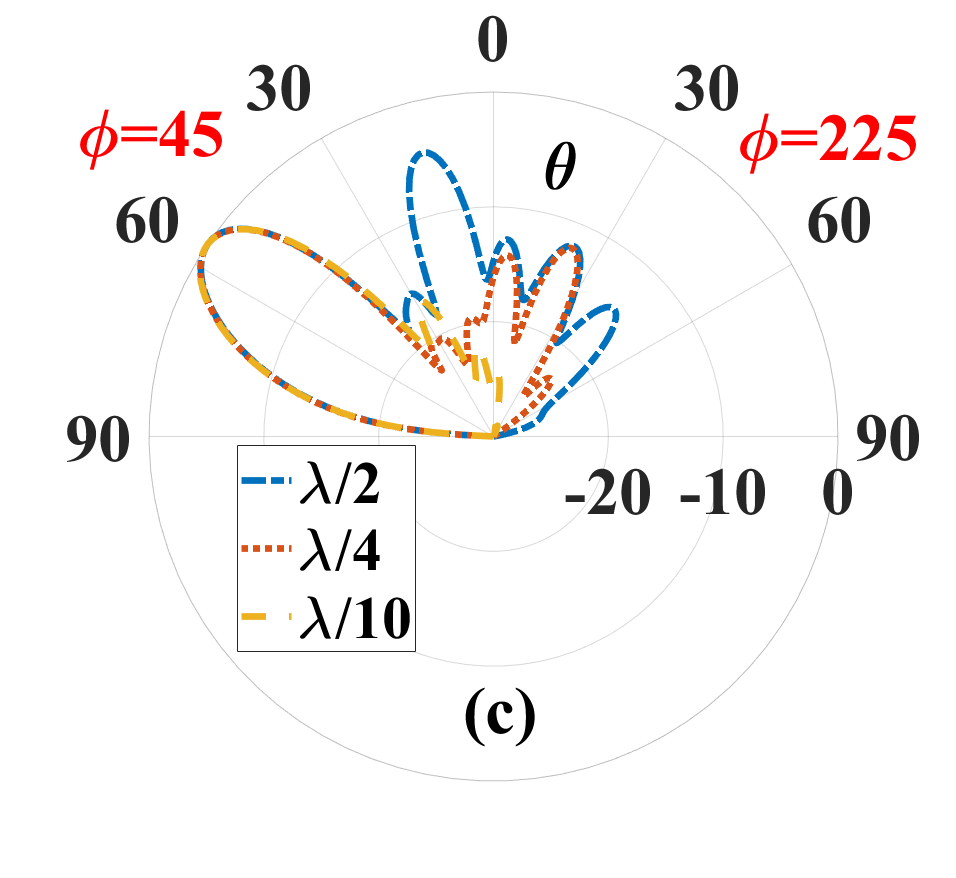}
    \includegraphics[width=0.49\columnwidth]{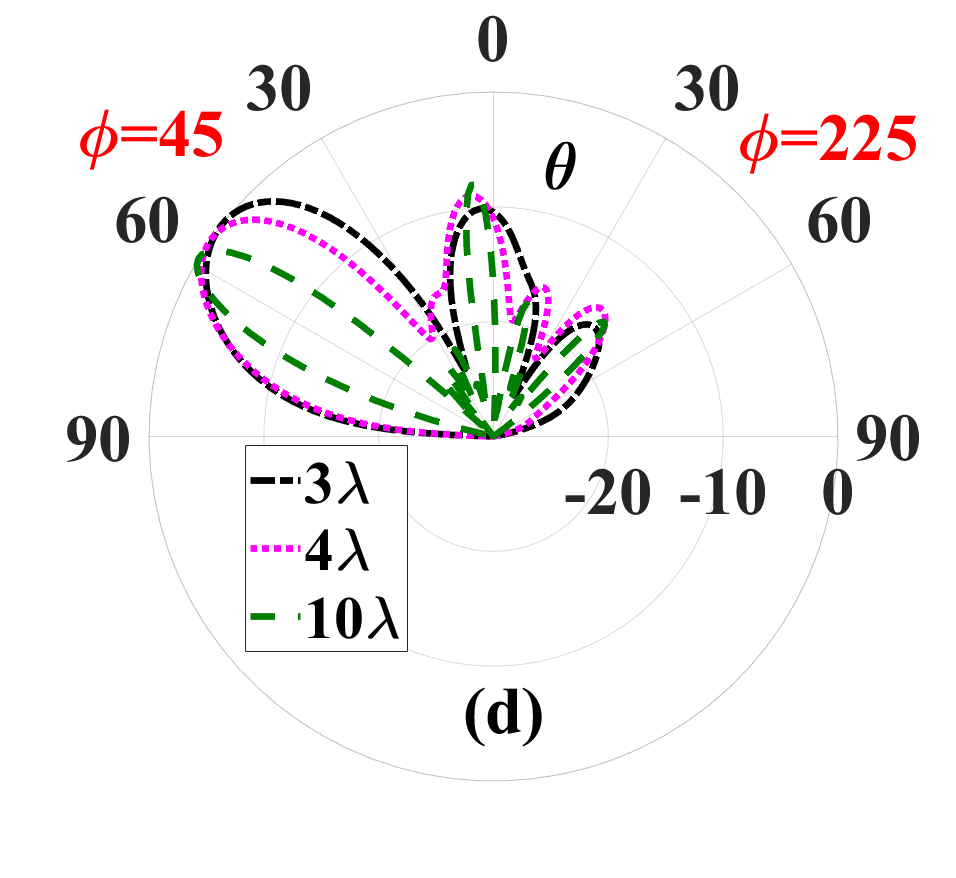}
    \vspace{-0.5cm}
    \caption{Normalized 2D/E-plane scattering patterns of ideal 4-state MS steering normally incident plane wave to $\theta_{r}=30^\circ$ (a--b) and $\theta_{r}=60^\circ$ (c--d). Panels (a) and (c) are for different unit cell sizes at fixed aperture $D_m = 5\lambda$, whereas panels (b) and (d) are for different apertures with fixed cell size $D_u = \lambda/3$.}
    \label{E-plane}
\end{figure}

We will start by assessing the effect of aperture and cell size on the performance of 4-state MS with ideal unit cell response for a few different scenarios. To this end, normalized 2D (E-plane) scattering patterns are presented in Fig.~\ref{E-plane}; the plots correspond to steering from normal incidence to two reference directions, namely $\theta_r=30^o$ and $60^o$, while $\phi_r=45^\circ$ in both cases. Moreover, we consider three cell sizes $D_u=\{\lambda/2, \lambda/4, \lambda/10\}$ for a fixed aperture $D_m=5\lambda$, and then three apertures $D_m=\{3\lambda, 4\lambda, 10\lambda\}$ for a fixed cell size $D_u = \lambda/3$. The resulting patterns clearly illustrate that targeting elevated angles (near zenith) leads to better results, due to their proximity to the specular reflection direction; in contrast, targeting ground-level (grazing) reflection angles, significant side lobes appear while the main lobe becomes wider, due to the `steeper' phase gradients applied across the MS. The results also re-iterate our previous conclusions on the effect of aperture and cell size, now confirmed for various reflection directions: higher apertures always improve (reduce) the HPBW whereas smaller cells always improve (reduce) the SLL. Note that the maximum directivity also increases with aperture (not shown in these normalized plots).

\begin{figure}[!t]
    \centering
    \hspace{-0.5cm}\includegraphics[width=0.46\columnwidth]{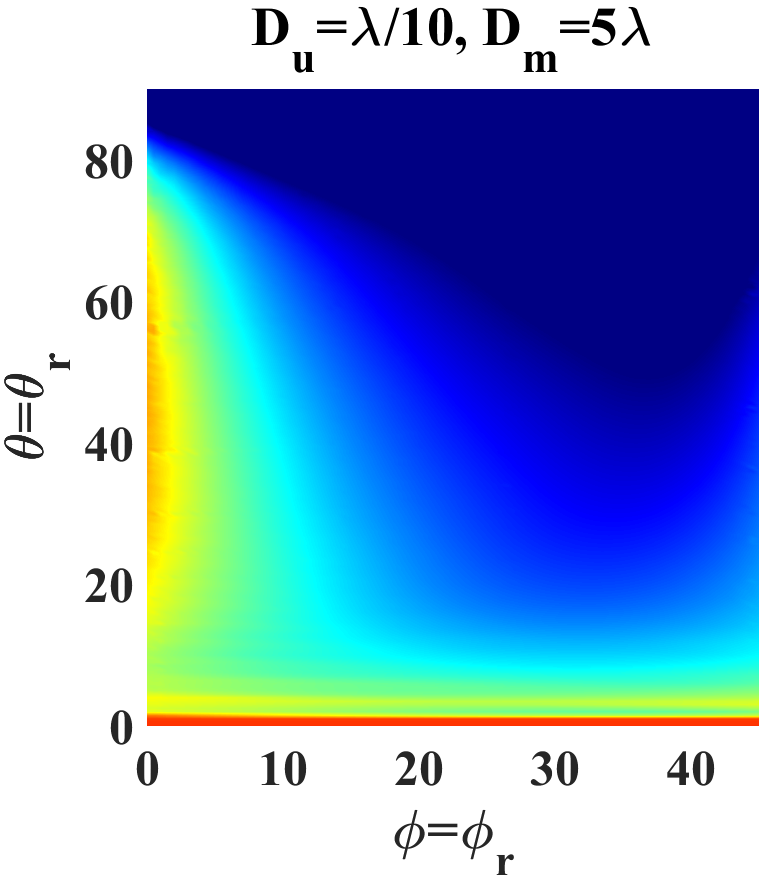}
    \includegraphics[width=0.46\columnwidth]{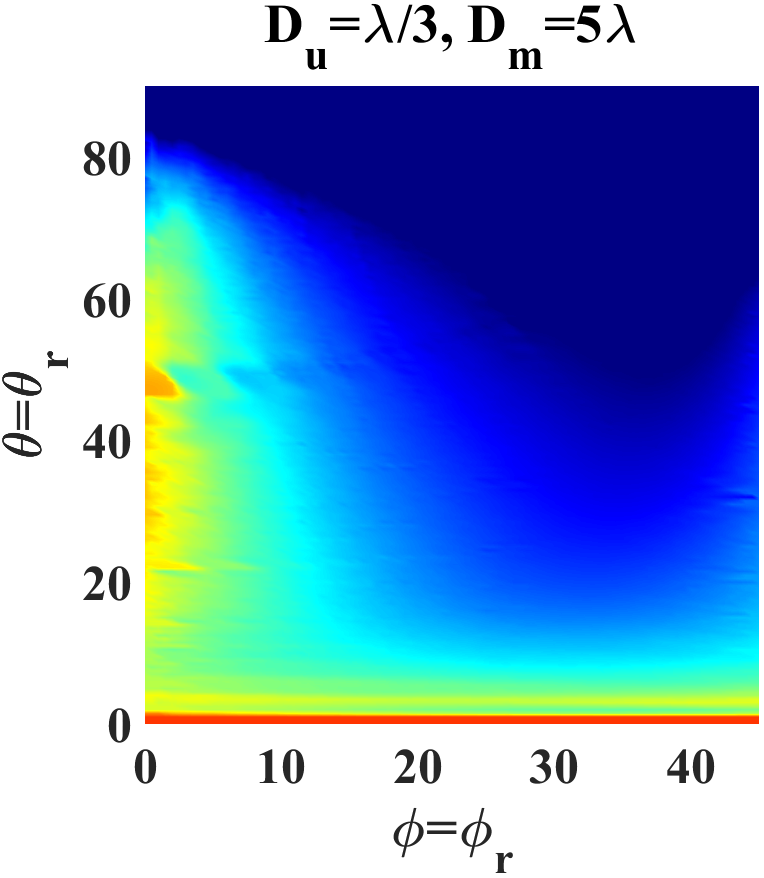}
    \includegraphics[width=0.46\columnwidth]{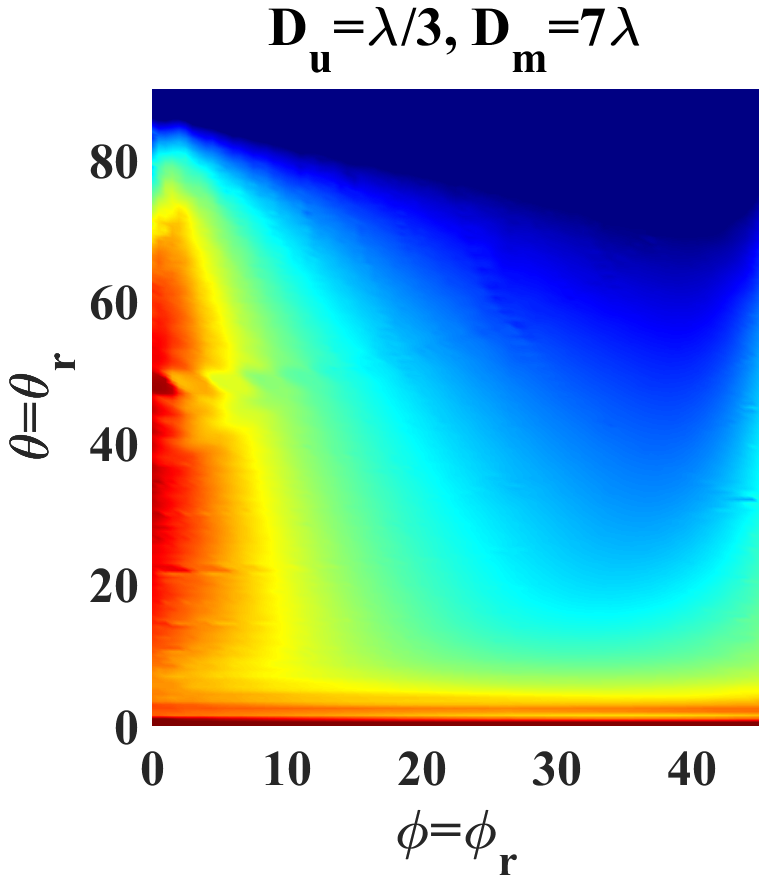}
    \includegraphics[width=0.51\columnwidth]{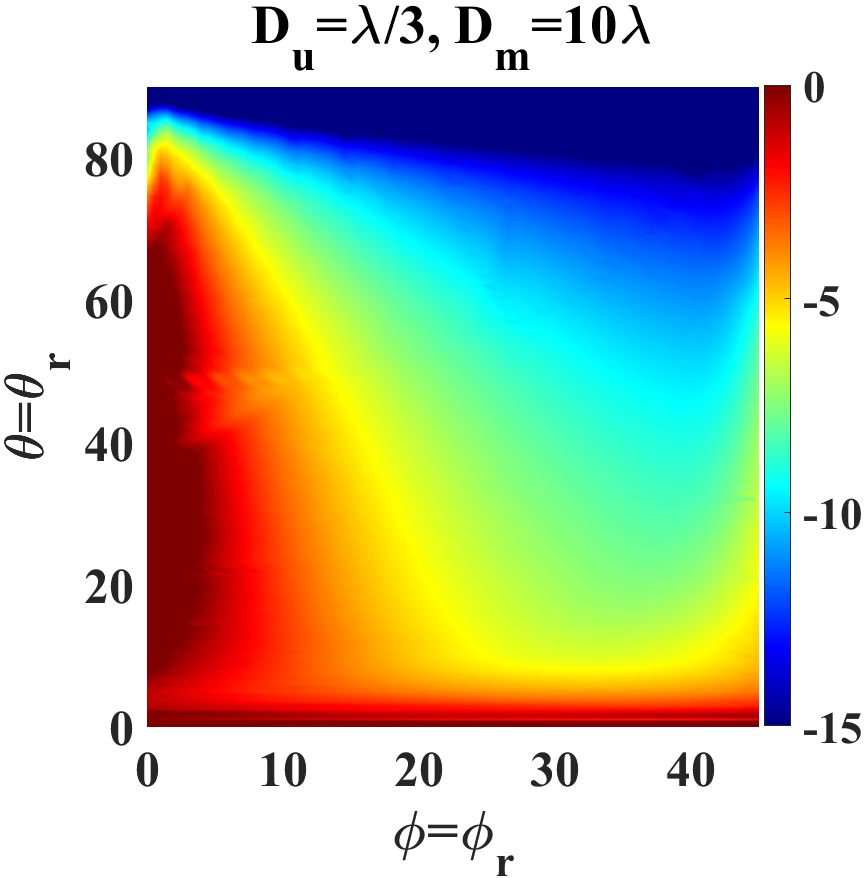}
    \vspace{-0.2cm}
    \caption{Normalized directivity when steering a normally incidence plane wave to any direction in the quarter-hemisphere. Each of the four panels corresponds to a different unit cell ($D_u$) and aperture size ($D_m$) combination. In all cases, we assume four ideal phase states, i.e., 2-bit encoding of the metasurface.}
    \vspace{-0.7cm}
    \label{Coverage}
\end{figure}

To generalize the example presented above, we repeat the analysis for multiple steering directions to the upper hemisphere and, in each case, evaluate the directivity as a representative performance metric. Hence, we extend previous works \cite{Wei:13, Forouzmand2016, 8668465} where only a set of discrete angles were studied, as the analysis of the complete angular space is extremely time-consuming unless analytical methods are used to focus on the scaling of the dimensional parameters instead. Figure \ref{Coverage} plots the normalized directivity when steering from normal incidence to: $\theta_r=0\rightarrow90^\circ$, and $\varphi_r=0\rightarrow45^\circ$. The region for which the normalized directivity is above a certain value is considered the coverage zone of the MS. Our analysis also amounts for variable cell and aperture size: $D_u=\lambda$/3, $\lambda$ /10 and $D_m=5\lambda$, $7\lambda$, $10\lambda$. We confirm that the performance is consistently better in directions close to the specular reflection (normal, in this case) and get worse as we approach steering directions close to the MS plane. The azimuth angle has a smaller influence on the performance. 

\subsection{Impact on Metasurfaces with Realistic Unit Cells}
\label{sec:Inc}
Here, we consider a fixed MS of $D_m=5\lambda$ aperture composed of the realistic unit cells designed in Section \ref{sec:unit cell}; these are 4~mm wide, i.e., $D_u\approx\lambda/3$ for $f\approx 25$~GHz. We consider wave incidence from three directions, $\theta_i=\{0,30^\circ,60^\circ\}$ and $\varphi_i=0$ in all cases. For this MS, we calculate the performance metrics as a function of the reflection direction requested, $\theta_r=0$ to $85^\circ$ and $\varphi_r=45^\circ$, after mapping the required phase-profile for each steering scenario onto the four available states. Note that the steering scenario that we selected corresponds to off-plane retro-reflection, which is more demanding compared to scenarios like in-plane steering or steering close to the specular reflection.

\begin{figure}[!t]
    \vspace{-0.4cm}
    \includegraphics[width=1\columnwidth]{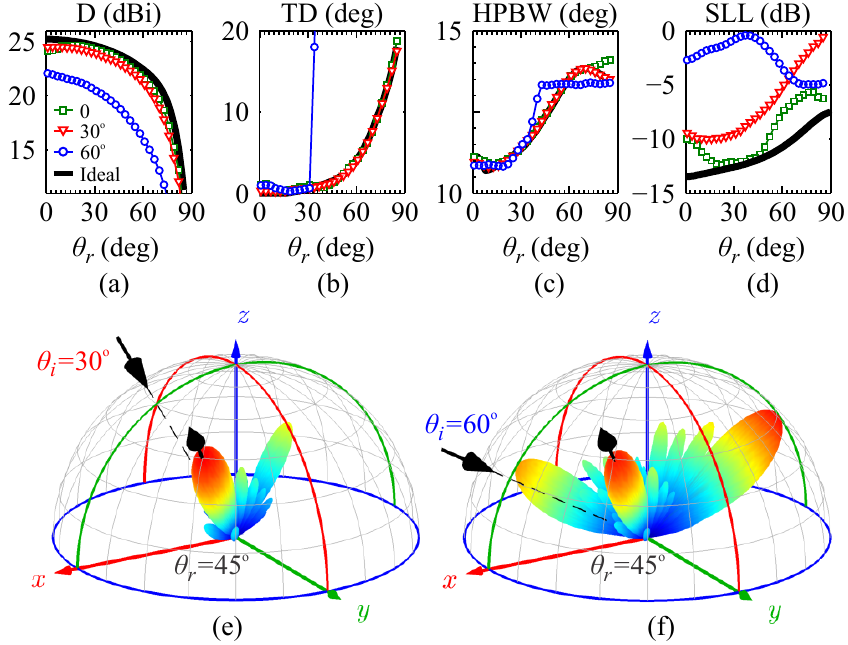} 
    \vspace{-0.4cm}
    \caption{Realistic metasurface performance metrics as a function of steering direction $(\theta_r,45^\circ)$ and three incidence directions, ($\theta_i$,0). (a) Directivity at $\theta_r$, (b) Target deviation, (c) HPBW, (d) SLL. The thick black curves correspond to the ideal case of continuous phase profiling of the MS. Logarithmic-scale 3D scattering patterns for two reference cases, (e) $\theta_i=30^\circ$ and (f) $\theta_i=60^\circ$, targeting steering to $\theta_r=45^\circ$ in both cases. The MS has $D_m=5\lambda$, $D_u=\lambda/3$ and its non-ideal states (amplitude and phase of reflection coefficients) are four, corresponding to four capacitance values.}
        \vspace{-0.4cm}
    \label{fig:RealCellsMetrics_sweepTheta}
\end{figure}

The resulting curves presented in Fig.~\ref{fig:RealCellsMetrics_sweepTheta}(a)-(d), including also the absolute limit values corresponding to ideal (continuous) phase profiling, indicate that the realistic unit cell design is capable of almost optimal performance for slightly oblique incidence, with respect to the directivity, HPBW and TD metrics; performance degrades with increasing $\theta_r$ (steering further away from specular direction) and $\theta_i$ (coming closer to grazing incidence), while the curves are generally monotonic and smooth. The notable exception is SLL which diverges from the ideal trendline even for the reference case of normal incidence; this is attributed firstly to the relatively large unit cell, secondarily to the `nearest neighbour' staircasing used to optimally map the continuous phase profile to the given fixed states for each steering direction, and, finally, to our post-processing algorithm which takes into account only the highest directivity side lobe, in whichever direction it might appear. For this fixed MS and demanding steering scenario, the performance breaks down for $\theta_i=60^\circ$ and $\theta_r>30^\circ$, due to the strong presence of a parasitic lobe in the specular direction; this can be visualized in Fig.~\ref{fig:RealCellsMetrics_sweepTheta}(e) and (f), depicting the scattering patterns acquired for slightly oblique and highly oblique incidence, respectively, when the steering direction is ($\theta_r,\varphi_r)=(45^\circ,45^\circ)$.

Increasing the pool size of the available phase states (capacitance values), from 4 to 8 or 16, would lead to progressively better performance, i.e., all metric curves would get closer to the ideal profile curves, even for highly oblique incidence. As discussed in Section \ref{sec:unit cell}, this improvement is due to the higher reflection-phase span (coverage) that can be attained with optimal selection of capacitances from a finer-resolution and/or wider pool. Finally, note that owing to the adopted unit cell design approach, the overall performance is better as the incidence angles decreases (closer to zenith), while TE polarization behaves better than TM; however, the unit cell can in principle be designed for any reference case, e.g., for TM polarization and/or for highly oblique incidence. 

\section{Discussion}
\label{sec:discussion}
This section aims to illustrate how the proposed methodology can be leveraged to guide the dimensioning of programmable MSs. Section \ref{sec:guidelines} discusses the extraction of design guidelines from the exploration, Section \ref{sec:fom} exemplifies the use of combined figures of merit to delimit the practicable design space, and Section \ref{sec:cost} describes how cost could be introduced in the exploration. 

\subsection{Extracting Design Guidelines from Performance} 
\label{sec:guidelines}
As expected, previous sections have confirmed that large metasurfaces with small discretization error (unit cell size tending to zero) and phase quantization error (large number of unit cell states) consistently yield the best performance for beam steering. 
However, the trends depend much on the performance metric and some metrics have clear \emph{frontiers} where performance increases abruptly. For instance, we have seen that, as expected, unit cell sizes below $\lambda/2$ are required to achieve reasonable directivities and side-lobe levels.

The scaling trends with respect to the number of unit cell states lead to less anticipated results. It has been observed that at least four states ($N_s=4$) are needed to achieve high-quality steering performance and that, while additional bits help in suppressing the side-lobe level and increasing the directivity, the improvements soon saturate. We have also seen that having a larger pool of available states is necessary to increase the angular range of the MS. In Fig.~\ref{AP}, we have shown that a pool of $4N_s$ states instead of $N_s$ states can perfectly accommodate incidence angles of $30$ and $60$ degrees for both polarizations.

\subsection{Application-Specific Figures of Merit} 
\label{sec:fom}
Thus far, the study has been application-agnostic in the sense that specific performance metric combinations are not taken into account. For instance, it is a well-known problem that, although narrow beams provide high efficiency and may be in fact necessary in some SDM/RIS-enabled scenarios \cite{Akyildiz2018}, slight target deviations can lead to loss of connectivity. Wider beams are less efficient, but also less prone to disruption.



The methodology presented in this paper can help reason about multiple design decisions, thereby delimiting the practicable design space, when putting different performance metrics together and introducing user requirements. For instance, beam steering for 5G communications will generally demand low beamwidth with low side-lobe level to minimize interference. Let us assume, as a practical case, a scenario where the necessary quality of experience is achieved with a $HPBW$ of 20 degrees with $\pm$5 degrees of tolerance and a $SLL$ of -13 dB with $\pm$2 dB of tolerance. In this context, we could define a figure of merit that encompasses both requirements. 
Although a formal definition of such a figure of merit is outside the scope of this work, we propose a particular example as follows
\begin{equation}
FoM_{1} = 1 - w\cdot \delta(HPBW) - (1-w)\cdot \delta(SLL)
\label{eq:FoM}
\end{equation}
where $w\in[0,1]$ is the weight of the $HPBW$ metric and $\delta(\cdot)$ is the distance of a metric to its nominal required value, normalized to the tolerance range. We set $FoM_{1} = 0$ if the design point is outside the tolerance interval.

Figure \ref{fig:perfFom} shows the $FoM_{1}$ for the conditions mentioned above for $N_s = 4$ and normal incidence. A value of 1 indicates maximum suitability of a design point, whereas a value of 0 delimits invalid design points. In this case, values around $D_m = 4\lambda$ for $D_u < 2\lambda/5$ are a good fit for the proposed application. Making an analogy to networking provisioning, one could argue that MSs with $D_m > 4\lambda$ and unit cells of lateral size $D_u < \lambda/3$ tend to be \emph{overprovisioned} as they perform better than the requirements set. whereas the MS is \emph{underprovisioned} for $D_m < 3\lambda$ or $D_u > 2\lambda/5$. Finally, note that while we considered that both metrics are equally important ($w=0.5$), architects can define their own weights depending on the application.

\subsection{Performance-Cost Analysis}
\label{sec:cost}
It has been shown throughout the paper that optimum performance is obtained in asymptotic cases of very large MSs with very small unit cells and a high number of states, which is clearly impractical. Although defining the application's requirement and tolerance interval helps to delimit the design space, practical design guidelines need to consider cost and complexity. To bridge this gap, parameterized models accounting for the cost or power consumption associated to integrated circuitry can be incorporated to our methodology for a joint performance-cost analysis. This would allow system architects to quantify the different tradeoffs with performance-cost figures of merit and, by adding weights to each metric, find the optimal design space for a particular budget.

\begin{figure}[!t]
\vspace{-0.4cm}
\subfigure[Performance figure of merit. \label{fig:perfFom}]{ \includegraphics[width=.46\columnwidth]{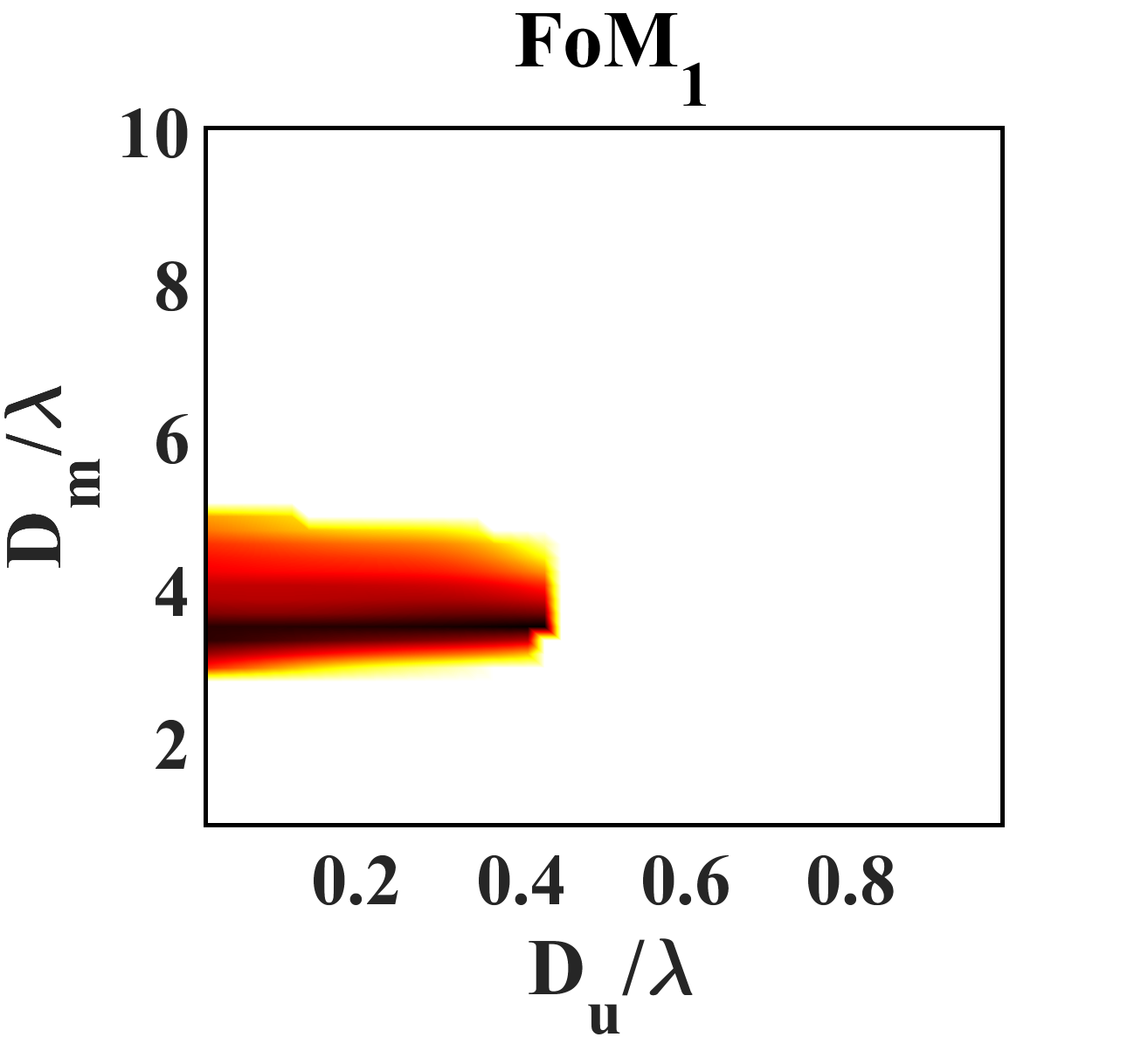} }
\subfigure[Performance-cost figure of merit. \label{fig:perfCostFom}]{ \includegraphics[width=.46\columnwidth]{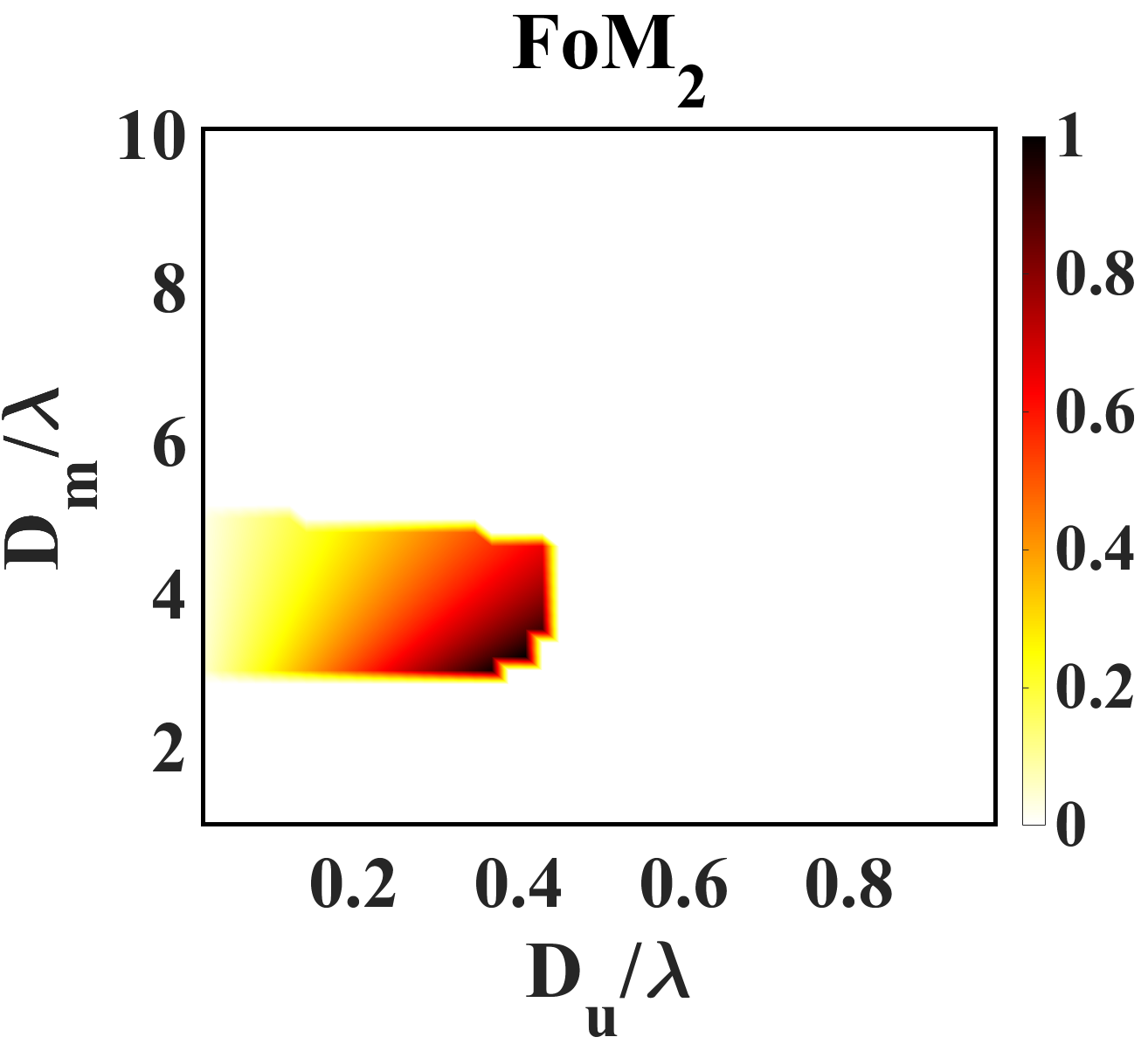} }
\caption{Evaluation, through figures of merit, of a 4-state MS for beam steering with a beamwidth requirement of $HPBW=20^{o} \pm 5^{o}$ and side-lobe level requirement $SLL=-13 \pm 2$ dB. Values close to 0 (1) refer to invalid (optimal) design points.}
\vspace{-0.4cm}
\label{fig:fom}
\end{figure}

To exemplify the process, let us consider the example from previous section and assume that power or cost of the MS scale linearly with the number of unit cells per dimension. This assumption is backed up by recent studies analyzing the impact of adding more controllers to the amount of internal messages required to reprogram the MS \cite{kouzapas2020towards, Saeed2019}. In our particular example, our performance-cost figure of merit is named $FoM_{2}$ and is obtained by dividing $FoM_{1}$ from Eq. \eqref{eq:FoM} by the number of unit cells per dimension and normalizing the result. As shown in Figure \ref{fig:perfCostFom}, the tendency is to favor configurations with less unit cells within the range that yields good performance within the tolerance range, as the intuition would suggest.

\section{Conclusion}
\label{sec:conclusions}
This paper has presented a methodology for the design-oriented scalability analysis of programmable metasurfaces (MSs), which allows to obtain a set of performance metrics across the design space. We have applied the methodology to analyze the beam steering case, evaluating the scaling trends of the directivity, target deviation, half power beam width, and side-lobe level with respect to multiple dimensional and programming parameters. We have observed that four unit cell states (2 bits) are enough to provide acceptable performance and confirmed that, as expected, large MSs with small unit cells provide the best performance. We further confirm that the performance drops significantly as incidence or target reflection angles approach the MS plane due to a degradation of the unit cell response. From the analysis, we conclude that the $\theta_r < 60$\textsuperscript{o} range is practicable for most MS designs and that, beyond that angle, increasing the amount of unit cell states may alleviate the performance degradation to some extent. Finally, we proposed the use of figures of merit that, tied to user requirements and cost models, provide an assessment of the practicable design space and optimal regions of such space in an attempt to guide the development of programmable MSs for future SDM/RIS-enabled wireless environments. 


\section*{Acknowledgment}
This work has been supported by the European Commission under grant H2020-FETOPEN-736876 (VISORSURF) and by ICREA under the ICREA Academia programme. Odysseas Tsilipakos acknowledges the financial support of the Stavros Niarchos Foundation within the framework of the project ARCHERS (‘‘Advancing Young Researchers Human Capital in Cutting Edge Technologies in the Preservation of Cultural Heritage and the Tackling of Societal Challenges’’).

\bibliographystyle{IEEEtran}

\end{document}